\documentstyle [12pt] {article}
\newcommand{\PD    }[2]{\frac{\partial#1}{\partial#2}}

\newcommand{\OP    }{\left(  }
\newcommand{\CL    }{\right) }

\begin{document}

\title{ \bf Global Phase Diagrams of Mixed Surfactant-Polymer
Systems at Interfaces}
\author{
Xavier Ch\^{a}tellier$^{(1,2,*)}$ and David Andelman$^{(1)}$
\and
$^{(1)}${\it School of Physics and Astronomy}\\
{\it Raymond and Beverly Sackler Faculty of Exact Sciences}\\
 {\it Tel Aviv University, Tel Aviv 69978 Israel}\\
\and
$^{(2)}${\it Department of Materials and Interfaces}\\
{\it Weizmann Institute of Science}\\
{\it Rehovot 76100 Israel}
}
\date{28 November 1995}
\maketitle

\begin{abstract}

Insoluble surfactant monolayers at the air/water interface undergo a
phase transition from a high-temperature homogeneous state to a
low-temperature demixed state, where dilute and dense phases coexist.
Alternatively,
the transition from a dilute phase to a dense one may be induced by
compressing the
monolayer at constant temperature.
We consider the case where the insoluble surfactant monolayer interacts with a
semi-dilute polymer solution solubilized in the water subphase.
The phase diagrams of the mixed surfactant/polymer system are investigated
within the framework of mean field theory. The polymer enhances the
fluctuations of the monolayer and induces an upward shift of the critical
temperature. The critical concentration is increased if the monomers are more
attracted (or at least less repelled) by the surfactant molecules than by the
bare water/air interface. In the case where the monomers are repelled by the
bare interface but attracted by the surfactant molecules (or vice versa),
the phase
diagram may have a triple point. The location of the polymer
special transition
line appears to have
a big effect on the phase diagram of the surfactant monolayer.

\end{abstract}

\vspace*{3cm}

\noindent
$^{(*)}$Present address:
{CRM--ICS, 6, rue Boussingault, 67083 Strasbourg Cedex, France}
\thispagestyle{empty}

\pagebreak

\setlength{\textheight}{23cm}

\section{Introduction}

Understanding the subtle interactions between macromolecules, such
as polymer or
proteins, and amphiphiles, such
as surfactants or phospholipids, has been a
problem of prime interest in recent years in many industrial applications
and in biological systems. For instance, biomembranes
\cite{Sackmann1,Bloom} are usually depicted as fluid bilayers composed of
different constituents: phospholipids, cholesterol and proteins. In
addition, a complex macromolecular network ({\it the cytoskeleton}) is
associated with the inner side of the bilayer and modifies the mechanical
properties of the membrane, while the glycocalix, on the outer side, is
believed to play an important role in molecular recognition. In industry,
surfactants are used in a wide range of applications (detergents, soaps,
oil recovery, paints), where polymers are often added in order to
provide stability for the system, especially in the case of colloidal
suspensions and oil/water emulsions \cite{Napper}. Those mixed polymer and
surfactant systems tend to create complex self-assembly structures
(connected micelles, gels, networks, etc.)
\cite{Goddard}--\cite{Iliopoulos}.
Finally, drug delivery via
micro-encapsulation is an other example where the stability of surfactant
vesicles is improved by the adsorption of polymer \cite{Benita}.

In recent years, a new category of {\it associating} polymers has been
introduced. Those are the hydrophobically modified water soluble polymers
(HM-WSP), consisting of a water soluble polymer backbone carrying small
hydrophobic side chains. Such polymers present interesting properties of
self-association, which may even be enhanced by the addition of
surfactant, and are very useful as viscosity modifiers of aqueous
solutions \cite{Iliopoulos,Wang}. The subtle coupling between the surfactant
and the polymer may lead to unusual phenomena like thermogelation
\cite{Olsson}, where gelation of the system is obtained upon increasing of
the temperature. Such systems have been studied theoretically \cite{Nikas}
as well as experimentally in the bulk.
However, little is known about their
behavior at interfaces \cite{Chari}.

In the present work, the interaction of water soluble polymers with a
surfactant monolayer located at the air/water interface is considered. We
restrict ourselves to the relatively simple situation of an insoluble
surfactant monomolecular layer ({\it Langmuir monolayer}). Langmuir
monolayers have been used in many applications \cite{Gaines,Ulman}, from
evaporation control to non-linear optic devices (via the creation of {\it
Langmuir-Blodgett monolayers}). They are also used to study
crystallization of solids \cite{Lahav} and provide useful model systems
for more complicated fluctuating liquid interfaces (membranes) where
curvature effects can not be neglected.

Another motivation for the present study comes from the lack of
understanding of adsorption (or depletion) of polymers close to non-ideal
interfaces, as compared with adsorption on ideal surfaces (namely,
perfectly flat and chemically homogeneous).
On ideal surfaces \cite{Fleer}--\cite{scalingconcepts}
theories for neutral and flexible polymers in good solvent have been
performed (both for adsorption and depletion) and compared with
scaling theories. For non-ideal surfaces, much less theoretical works
exists. It has been suggested that the bending properties of a curved
interface are modified when in presence of adsorbing polymer
\cite{Marques}--\cite{deGennes2}. When the polymer adsorbs on both sides of
the interface (a bilayer for instance), the curvature modulus decreases,
while the saddle-splay modulus increases. When it adsorbs only on one
side, a non-zero spontaneous curvature is induced. The situation of a
perfectly flat but chemically heterogeneous interface has been considered
only in a few works \cite{deGennes2}--\cite{Xavier}. The case of
annealed disorder (namely, when the disorder is at thermodynamic
equilibrium and the heterogeneities can
diffuse laterally) is found to behave differently
from the case of quenched disorder (where
heterogeneities are spatially ``frozen").
However, in both cases the adsorption
of the polymer is increased by the non-ideality of the surface.
In this context, a
surfactant monolayer is an example of a non-ideal annealed surface, where
the order parameter is the local surface surfactant concentration.

The phase diagram of surfactant monolayers can be constructed as a
function of the thermodynamical variables \cite{Gaines}: surface pressure
and temperature (or equivalently area per molecule and temperature). At
low surface pressure, a phase separation occurs (for temperatures below
the corresponding critical temperature). Dilute (gaseous) and dense
(liquid-expanded) regions of the monolayer coexist, in analogy to phase
transitions in the bulk. In the phase diagram, single-phase and two-phase
regions are separated by a coexistence curve. At higher surface pressure,
other phase transitions occur. Depending on the symmetries of the specific
surfactant molecules, the phase diagrams
 are more complex and still a topic of
current investigation \cite{Knobler,Kaganer}.

In the following, we consider how a simple condensation
transition (gas to liquid expanded)
of a surfactant monolayer at the air/water interface
is affected by the presence of polymer in the water subphase. The free
energy and the assumptions used in deriving it are introduced in Sec. 2,
while in Sec. 3 we discuss the main results, as applied to a simple case;
the general theory is detailed in Appendices A and B. Finally, some
analytical considerations on the critical point are presented in Appendix C.

\section{The polymer/surfactant free energy}

The model used for the mixed surfactant/polymer system follows closely the
lattice model introduced in Ref. \cite{AndJoa}. The local (dimensionless)
free energy per site $F$, rescaled in units of $k_{B} T$, $k_B$ is the
Boltzmann constant and $T$ is the temperature, can be separated into three
parts: the surfactant contribution $F_{s}$, the polymer contribution
$F_{p}$, and the coupling term $F_{ps}$:
\begin{equation}
 F = F_{s} + F_{p} + F_{ps}
\label{eq:general}
\end{equation}
In the following, those three terms are discussed separately.

\subsection{The surfactant contribution $F_{s}$}
The monolayer free
energy is calculated using a lattice-gas model. Each lattice site is
occupied either by a surfactant molecule or by an artificial vacancy, in
order to allow us to consider a compressible monolayer. The free energy of
a surfactant monolayer is the sum of the enthalpy and entropy of mixing
and depends on the monolayer area fraction (or equivalently coverage) $c$
ranging from zero to one, $c=A_{0}/A$, where $A_{0}$ is the close-packing
area of a surfactant molecule (or the area of one site on the lattice) and
$A$ is the actual area per surfactant molecule on the interface.
Typically
$A_{0}\simeq 25-35$  \AA$^{2}$
for a surfactant molecule \cite{Ben-Shaul}.
Disregarding linear terms, the surfactant free energy $F_{s}$ (per site
and per $k_{B}T$), within a Bragg-Williams (mean field) theory, is written
as:
\begin{equation}
   F_{s} = \nu^{-1} c \OP 1-c \CL +
c \log c + \OP 1-c \CL \log \OP 1-c \CL
\label{eq:surfactant}
\end{equation}
where $\nu^{-1}$ is the dimensionless interaction parameter of the
surfactant on the surface and describes Van der Waals interactions between
neighboring particles. The interactions between the
head groups of the surfactant molecules, playing an important role in the
determination of the highly compressed phases, as well as the freedom of
conformation for the hydrophobic chains, whose coupling with the
surfactant average $c$ is determinant in the liquid expanded versus liquid
condensed transition \cite{Ben-Shaul}, are not taken into account. As only
short-ranged interactions are considered (between neighboring sites), the
surfactant molecules are supposed to be neutral. The main interaction
modeled by the parameter $\nu$ are the Van der Waals interactions.

For an insoluble monolayer, the total number of surfactant molecules is
fixed. At low (and positive) values of $\nu$ (corresponding to low
temperatures), a phase separation between dense and dilute regions follows
from eq.~(\ref{eq:surfactant}). The stability of such a monolayer is
obtained by studying the convexity of the free energy ~\cite{Safran}. The
monolayer becomes unstable if the second derivative of the free energy
becomes negative. The condition $F_{s}''(c) = 0$ defines the {\it spinodal
line}, separating metastable and unstable regions. The spinodal line
obtained from eq.~(\ref{eq:surfactant}) is $\nu_{s}^{0}(c) = 2c (1-c)$
and it lies within the coexistence region of the phase diagram. In
addition, the coexisting curve limiting the two phase region ({\it the
binodal line}) is easily found from eq.~(\ref{eq:surfactant}) as the
system is symmetric about $c=0.5$:
\begin{equation}
   \nu_{b}^{0}(c)=- \frac{1-2c}{\log c - \log (1-c)}
\label{eq:spinsurf}
\end{equation}
The spinodal and binodal lines join together at the critical point $c=0.5$,
$\nu_{c}=0.5$. In Fig.~1 the binodal line and the critical point are shown
for a pure surfactant monolayer.

\subsection{The polymer contribution $F_{p}$}

The polymer in the sub-phase is assumed to be neutral and flexible as well
as in good solvent conditions, hence with a positive second virial
coefficient and no polymer-solvent phase separation. For a
semi-dilute polymer solution, a mean field theory applied to the Edwards
density functional method is commonly used \cite{Edwards,scalingconcepts}.
The free energy density is conveniently expressed as a function of the
variable $\phi (z)$ related to $c_{p}(z)$, the local monomer
concentration, by $\phi^{2} (z)=c_{p}(z)/c_{b}$. The coordinate $z$
denotes the perpendicular distance from the interface, and
$c_{b}=c_{p}(z\rightarrow \infty )$ is the concentration of the polymer in
the bulk (acting as a reservoir). The characteristic length in the
solution is the Edwards correlation length $\xi=a/ \sqrt{3 v c_{b}}$,
where $v$ is the excluded volume parameter (positive, in good solvent
conditions), and the typical energy parameter (per $k_{B} T$) for the
interactions between monomers in the bulk is $\epsilon_{p}= A_{0} \xi
v c_{b}^{2}$. Using these notations, the free energy per site reads:
\begin{equation}
   F_{p}= \frac{\epsilon_{p}}{2} \int _{0}^{\infty} \,dz \OP\xi
   (\nabla \phi )^{2} + \frac{1}{\xi} (\phi ^{2} - 1)^{2} \CL
\label{eq:polymer0}
\end{equation}

 The first term accounts for the elastic flexibility of the polymer chains
and the second originates from the excluded volume interaction combined
with the equilibrium condition with the polymer bulk reservoir. The
polymer free energy $F_{p}$ is a functional of the polymer profile ${\phi
(z)}$ and of the order parameter at the interface $\phi_{s}=\phi (z=0)$.
It does not include the energy of interaction with the surface, discussed
separately below.

Minimizing $F_{p}$ with respect to the polymer profile ${\phi (z)}$,
leaving the surface value as a free parameter, yields the polymer
profile $\phi (z) =\coth (z/\xi+b)$, where $b$ is a constant of
integration related to $\phi_{s}$ by $\phi_{s} = \coth b$, in the case of
adsorption ($\phi (z)>1$), and $\phi (z)= \tanh (z/\xi +b')$ in the case
of depletion ($\phi (z)<1$), where, similarly, $\phi_{s} = \tanh b'$. For
both adsorption and depletion, the free energy $F_{p}$ for the optimal
profile is
\begin{equation}
   F_{p}= \frac{\epsilon_{p}}{3} \OP \phi_{s}^{3} - 3 \phi_{s} + 2 \CL
        = \frac{\epsilon_{p}}{3} \OP \phi_{s} -1 \CL ^{2} \OP \phi_{s}+2 \CL
\label{eq:polymer}
\end{equation}
and has a minimum at $\phi_{s}=1$. This means that the polymer solution
would like to be homogeneous throughout the solution at the imposed bulk
value $c_{p}(z)=c_{b}$. The only possibility of obtaining a profile with
$\phi_{s}\neq 1$ is due to the short range coupling of the polymer with
the surface. This coupling includes the surfactant monolayer as well as
the bare air/water interface. It is given below by the term $F_{ps}$.

A quantity accessible to experiment which measures the total adsorption of
the monomers at the interface is the polymer surface excess defined as
$\Gamma = \int_{0}^{\infty} \,dz (c_{p}(z) - c_{b})$. Using the above
results of the minimization for the polymer profile
(mean-field theory), it is simply related to $\phi_{s}$ by

\begin{equation}
\Gamma = c_{b} \xi (\phi_{s}-1)
\label{eq:surfexc}
\end{equation}

Note that, if $\rho$ is the volume fraction of the monomers in the bulk
solution, a naive calculation starting from $c_{b} \simeq \rho/a^{3}$
(where $a$ is the size of a monomer) yields $\epsilon_{p} \simeq
\rho^{3/2}$. For a semi-dilute polymer solution, $N^{-4/5}\ll \rho \ll 1$
(where $N$ is the number of monomers in a chain). Hence, roughly, for
$N=10^{4}$ the range for typical values of $\epsilon_{p}$ is given
\cite{Sam} by
$10^{-4}<\epsilon_{p}<10^{-1}$.

Although the self-consistent field theory provides a convenient and
qualitatively correct framework for the description of the semi-dilute
polymer solution, some of its predictions (like the form of the polymer
profile $c_{p}(z)$ for instance) are in disagreement with a scaling theory
\cite{scalingconcepts}. Nevertheless, we will use it to model the polymer
behavior in solution.


\subsection{The coupling term $F_{ps}$}
A bilinear term in the surfactant and monomer concentrations at the
interface ($z=0$) is a simple, yet meaningful, phenomenological coupling
for the polymer-interface interaction, which is
assumed to be short-ranged:
\begin{equation}
   F_{ps} = - \frac{1}{2} [\alpha_{0} c + \gamma_{0} (1-c)] \phi_{s}^{2}
          = - \frac{1}{2} \epsilon_{ps} (c-c^{\star}) \phi_{s}^{2}
\label{eq:coupling}
\end{equation}
$\alpha_{0}$ is the polymer/surfactant interaction parameter, and
$\gamma_{0}$ is the polymer/bare interface interaction parameter. In
eq.~(\ref{eq:coupling}), we define the ``{\it effective}"
polymer/surfactant interaction parameter
$\epsilon_{ps}\equiv\alpha_{0}-\gamma_{0}$. It is positive whenever the
monomers interact more favorably with the surfactant molecules than with
the bare water/air interface. The {\it special transition} coverage is
defined as $c^{\star}\equiv -\gamma_{0}/ (\alpha_{0} -\gamma_{0})$. In
principle, those parameters can depend on temperature.

The phenomenological coupling $F_{ps}$ can be justified for polymers in
the semi-dilute regime since in such systems the monomers concentration is
small with respect to unity. However, it represents only the lowest term
in an expansion in the surfactant concentration at the interface.
When $\epsilon_{ps}>0$, the $F_{ps}$ term corresponds to a repulsion
of the
polymer from the surface (depletion) for $c< c^{\star}$ and
to an attraction to the surface
(adsorption) for $c > c^{\star}$.  The special
transition line $c = c^{\star}$ occurs for physical values of the
coverage, $0<c^{\star}<1$, when the attraction (repulsion) of the monomers
with the surfactant molecules is in competition with the repulsion
(attraction) with the bare interface. In the $0<c^{\star}<1$ range,
a positive $\alpha_{0}$ is equivalent to having a positive $\epsilon_{ps}$
and means that the interaction between the monomers and the
surfactant molecules is attractive.


\subsection{The total free energy $F$}

Combining all three contributions, $F_{s}+F_{p}+F_{ps}$, we obtain the
total free energy (per site of the interface and per $k_{B}T$):
 \begin{eqnarray}
   F &=& \nu^{-1} c \OP 1-c \CL + c \log c
+ \OP 1-c \CL\log \OP 1-c \CL \nonumber \\&& + \frac{1}{3} \epsilon_{p} \OP
\phi_{s} - 1 \CL ^{2} \OP \phi_{s} + 2 \CL - \frac{1}{2}
\epsilon_{ps}(c - c^{\star}) \phi_{s}^{2}
\label{eq:total0}
\end{eqnarray}
Note that the energy is invariant under the transformation $\epsilon_{ps}
\rightarrow -\epsilon_{ps}$, $c \rightarrow 1-c$ and $c^{\star}
\rightarrow 1-c^{\star}$. Therefore, it will be assumed in the following
that $\epsilon_{ps}>0$ without loss of generality.

 The free energy $F$ is a function of $\phi_{s}$ and $c$. Minimizing it
first with respect to the polymer surface order parameter $\phi_{s}$ (mean
field approximation), we obtain
 \begin{equation}
\phi_{s}=\frac{\lambda}{2
\epsilon_{p}}+\sqrt{\OP\frac{\lambda}{2\epsilon_{p}}\CL^{2}+1}
\label{eq:phi}
\end{equation}
 where $\lambda = \epsilon_{ps} \OP c - c^{\star} \CL$ measures the
strength of the interaction between the polymer and the overall interface
(including the bare interface as well as the surfactant). Equation
(\ref{eq:phi}) relates $\phi_{s}^{2}(c)$, the concentration of monomers at
the interface, with the surfactant area fraction $c$.
Consequently, the entire polymer
profile and the polymer surface excess can be found as a
function of the surfactant concentration on the interface.

The limit of a very strong adsorption, $\lambda/\epsilon_{p} \rightarrow
\infty$ ({\it e.g.}, $c^{\star}\ll 0$), yields $\phi_{s} \simeq
\lambda/\epsilon_{p}\gg 1$. On the other hand, the limit of very strong
depletion, $\lambda/\epsilon_{p} \rightarrow -\infty$ ({\it e.g.},
$c^{\star}\gg 1$), yields $\phi_{s} \simeq |\epsilon_{p}/\lambda| \ll 1$.
For $\lambda/\epsilon_{p} \rightarrow 0$ ({\it e.g.}, $c \rightarrow
c^{\star}$), $\phi_{s}\rightarrow 1$, which means that the polymer
solution remains homogeneous: $c_{p}(z)=c_{b}$, even at the surface. As is
shown in Fig.~2a, the special transition line $c=c^{\star}$ divides the
parameter range into an adsorption region ($c>c^{\star}$) and a depletion
one ($c<c^{\star}$).

 If the surfactant monolayer is in the two-phase region, dilute and
condensed regions of the surfactant coexist, and the polymer adsorbs
differently on those regions because of its different affinity as
described by the parameter $\epsilon_{ps}> 0$. Note that as the curve
$\phi_{s}(c)$ is convex (see Fig.~2b), the polymer surface excess is
enhanced when the surfactant monolayer undergoes a phase separation.
Qualitatively, the convexity of the curve $\phi_{s}(c)$ indicates
\cite{AndJoa} that the surfactant concentration fluctuations increase the
average
polymer surface concentration $\phi_{s}$ and, hence, the polymer surface
excess $\Gamma \sim \phi_{s}-1$.

Equation (\ref{eq:phi}) shows how the surfactant molecules affect the
polymer adsorption. The main remaining task is to understand how the
polymer itself affects the phase diagram of the surfactant monolayer,
since the two problems are coupled. Using eq.~(\ref{eq:phi}),
$\phi_{s}=\phi_{s}(c)$, the total free energy can be written only as a
function of the surfactant coverage $c$:
\begin{equation}
   F(c)=    F_{s}(c) -  \frac{\epsilon_{p}}{6}\,  (\phi_{s}^{2} + 3) \phi_{s}
\label{eq:energy}
\end{equation}
 The study of the convexity of the total free energy $F(c)$ as a function
of $c$ (the only remaining
order parameter) determines the location of the {\it
modified} spinodal line. Similarly, the full phase diagram can be obtained
numerically from a common-tangent construction of $F(c)$.

For sake of clarity, the main physical features can be illustrated in a
simple situation. This is done in the next section, where $\epsilon_{p}$,
$c^{\star}$ and $\nu \epsilon_{ps}$ are taken to be independent of the
temperature $T$, and $\nu \sim T$. A more general treatment without any
assumptions on the $T$-dependence of the interaction parameters is
presented in Appendices A and B.

\section{Results}
\setcounter{equation}{0}

\subsection{$T$-dependence of the interaction parameters}
 $T/\nu$ is independent of the temperature if we assume that the
interaction potential of the surfactant molecules has an infinite
repulsive core followed by a weak attraction independent of the
temperature, $0<w\ll k_B T$. In such a case, the surfactant second virial
coefficient is given by $v_{s}=1-w/(k_B T)$. Expanding the energy $F_{s}$,
eq.~(\ref{eq:surfactant}), in powers of $c$ shows that the virial
expansion \cite{Safran} is equivalent to the Bragg-Williams theory (at low
concentrations) with $\nu^{-1}=w/(k_B T)$. Thus $k_B T/ \nu$ is
independent of the temperature. Note that the strength of this
attraction $w$ is related to the critical temperature of the pure
surfactant monolayer by $w= \nu^{-1}_{c} k_B T_c = 2 k_B T_c$.

For the semi-dilute polymer solution, taking only excluded volume
interactions between monomers and assuming an athermal solvent, $v$ and
$\epsilon_{p}$ are independent of $T$.

We also assume that the interaction between the monomers and the bare
interface is independent of the temperature, resulting in $\gamma_{0}\sim
1/T$ since the dimensionless $\gamma_{0}$ is rescaled in units of
$k_{B}T$. In a similar manner, neglecting the steric effects
between
the monomers and the surfactant molecules, and assuming that the
monomer/surfactant interactions are attractive, weak and short-ranged,
results in $\alpha_{0}\sim1/T$. Under these conditions $\nu \epsilon_{ps}$
and $c^{\star}$ are independent of $T$.

In conclusion, the phase diagram in the simplified case can be plotted in
the $(\nu,c)$ plane and is a cut of the {\it global} phase diagram
(presented in the appendices) plotted in the ($\nu,
c,\epsilon_{p},\nu\epsilon_{ps},c^{\star}$) space. The next subsections
will present some features of this simplified case.


\subsection{The ($\nu,c$) phase diagram}
 We limit ourselves to $\nu>0$, $\epsilon_{p}>0$ and $\epsilon_{ps}>0$;
$\nu>0$ corresponds to an attraction between the surfactant molecules;
$\epsilon_{p}>0$ follows from the assumption that the polymer is in a good
solvent (its excluded volume parameter $v$ is positive), and
$\epsilon_{ps}>0$ can be used without loss of generality, as was explained
in Sec. 2.4. The spinodal line $\nu_{s}(c)$ of the mixed surfactant-polymer
system is obtained from the condition
 \begin{equation}
   \PD{^{2}F}{c^{2}}=-2\nu^{-1} + \frac{1}{c(1-c)}-
\frac{\epsilon_{ps}^{2}}{\epsilon_{p}}\frac{\phi_{s}^{3}(c)}{\phi_{s}^{2}(c)
+1} =0
\label{eq:spin0}
\end{equation}
 The critical point is the extremum point on the spinodal, satisfying in
addition
\begin{equation} \PD{^{3}F}{c^{3}}=0 \label{eq:crit0}
\end{equation}
 Equation (\ref{eq:spin0}) shows that the spinodal temperature $\nu_{s}$
is shifted upwards \cite{AndJoa} and the region of instability is
increased. Physically, this general effect comes from the indirect
attractive interaction between the surfactant molecules induced by the
polymer, and was already explained elsewhere \cite{deGennes2}. Here, it is
represented by the term $-\epsilon_{p}(\phi_{s}^{3}+3\phi_{s})/6$ in the
free energy. It is bigger for larger values of the surfactant
concentration $c$ because $\phi_{s}(c)$ is an increasing function of $c$.
Consequently, the phase diagram is no longer symmetric around $c=1/2$. The
shift on the spinodal and binodal lines is bigger for the large values of
$c$ and the critical point is shifted to a concentration $c_{c}>1/2$.

In the following, we discuss several limits and try to show that (except
in the low-coupling limit) the position of the special transition line
$c=c^{\star}$ in the $(\nu,c)$ plane crucially affects the phase diagram.
This can be checked easily in the limits of very high and very low
surfactant concentration, where the scaling of the spinodal temperature
$\nu_{s}$ is analytically derived and depends on the relative position of
the spinodal line with respect to the special transition line.


\subsection{The limits
$c\rightarrow 0$ and $c\rightarrow 1$}
 Three cases can be distinguished in the $c \rightarrow 0$ limit of the
spinodal line:
\begin{itemize}
\item
 if $c^{\star}>0$, the polymer is strongly depleted from the interface.
The shift of the spinodal temperature $\nu_{s}(c)$ from the pure value
$\nu_{s}^{0}(c)$ is small and scales as $\nu_{s}-\nu_{s}^{0} \sim c^{3}$,
where $\nu_{s}^{0}=2 c(1-c)$ is the pure surfactant spinodal line (no
added polymer).
\item
On the other hand, if $c^{\star}<0$, the polymer is strongly attracted by
the interface and $\nu_{s}$ scales as $\nu_{s} \sim c^{1/3}$.
\item
 In the special case when $c^{\star}=0$, the polymer solution remains
homogeneous. The dominant term in the spinodal equation also changes and
$\nu_{s}$ scales as $\nu_{s} \sim c^{1/2}$. \end{itemize} The scaling of
the spinodal temperature in the limit $c\rightarrow 1$ depends similarly
on the position of the special transition line relatively to the line
$c=1$. In the following, the other regions of the phase diagram are
considered and $\nu$ is at least of order unity.

\subsection{The low-coupling limit: $\nu \epsilon_{ps}\ll 1$}
 In the case when $\nu\epsilon_{ps}$ is small enough \cite{smalleps}, the
third term in the spinodal equation (\ref{eq:spin0}) is negligible. The
phase diagram is very similar to that of the pure surfactant monolayer
(see Fig.~1). The spinodal temperature $\nu_{s}(c)$ as well as the binodal
temperature $\nu_{b}(c)$ can be expanded in powers of $\nu\epsilon_{ps}$.
To second order in $\nu\epsilon_{ps}$, the shift of the binodal line is
identical to the shift of the spinodal line:
 \begin{equation}
   \nu_{s}(c)-\nu_{s}^{0}(c)=
   \nu_{b}(c)-\nu_{b}^{0}(c)=
   \frac{1}{4\epsilon_{p}}(\nu\epsilon_{ps})^{2} \label{eq:lowexp}
\end{equation}
For the low-coupling limit, the critical temperature is
shifted upwards \cite{AndJoa} by a factor of order
$(\nu\epsilon_{ps})^{2}$. Note that, to second order in
$\nu\epsilon_{ps}$, the shift is independent on the surfactant area
fraction and the special transition coverage $c^{\star}$. The critical
concentration $c_{c}$ is also shifted upwards, but only to third order in
$\nu\epsilon_{ps}$, :
\begin{equation}
c_{c}-1/2=\frac{1}{8\epsilon_{p}^{2}} (\nu\epsilon_{ps})^{3}
\label{eq:conccrit}
\end{equation}
 Another case where the third term in the spinodal equation
(\ref{eq:spin0}) is negligible is when $\phi_{s}(c)\ll 1$. Here, the
polymer is very strongly depleted from the interface. This occurs, for
instance, when $\epsilon_{p}\ll 1$ and $c<c^{\star}$.

\subsection{The strong coupling limit: large values of $\nu \epsilon_{ps}$}
 We define the strong coupling limit as the situation when the spinodal
line is strongly shifted upwards, $\nu_{s}(c)\gg 1$ and the first term in
the spinodal equation (\ref{eq:spin0}) is negligible. $\epsilon_{ps}$ is
the only parameter left in the equation depending on the temperature. As
$\nu=(\nu\epsilon_{ps})/\epsilon_{ps}$, the temperature on the spinodal
line is proportional to the parameter $\nu\epsilon_{ps}$ and the
coefficient of proportionality $\epsilon_{ps}^{-1}$ is obtained from the
spinodal equation (\ref{eq:spin0}) rewritten as a fifth order equation for
$\epsilon_{ps}$ depending on $\epsilon_{p}$, $c^{\star}$ and $c$ (and
independent of $\nu\epsilon_{ps}$):
 \begin{equation}
   4\left(\frac{\epsilon_{p}}{c(1-c)}\right)^2 +\left(
\frac{c-c^{\star}}{c(1-c)}
\right)^2 \epsilon_{ps}^{2} - 4  \frac{c-c^{\star}}{c(1-c)}
\epsilon_{ps}^{3} -\epsilon_{ps}^{4} - \frac{(c-c^{\star})^{3}}
{\epsilon_{p}^{2}c(1-c)} \epsilon_{ps}^{5} =0
\label{eq:strong}
\end{equation}
 Whenever eq.~(\ref{eq:strong}) has a unique positive solution
$\epsilon_{ps}$, the approximation of the strong coupling limit is
self-consistent provided that $\nu\epsilon_{ps} \gg \epsilon_{ps}$. This
is the case for $c \geq c^{\star}$. On the other hand, when the polymer is
depleted from the interface (corresponding to $c<c^{\star}$), there is a
minimal value of $c$ for which eq.~(\ref{eq:strong}) has a positive
solution. Indeed, a sufficient condition for eq.~(\ref{eq:strong}) not to
have any positive solution is:
\begin{equation}
c^{\star}-c\,\,> \,\,\left( \frac{11}{1350 + 210 \sqrt{42}}\right) ^{1/4}
\sqrt{\epsilon_{p}}
\,\,\simeq\,\, 0.2524 \sqrt{\epsilon_{p}}
\label{eq:condit}
\end{equation}
 This implies that, in a system where the condition (\ref{eq:condit}) is
respected for the largest physical $c$ value, $c=1$, the strong coupling
limit can not be defined and the critical temperature $\nu_{c}$ is
necessarily of order unity.

When $c^{\star}\rightarrow -\infty$ or when $\epsilon_{p} \rightarrow 0$
(and $c>c^{\star}$), the solution of eq.~(\ref{eq:strong}) is small,
$\epsilon_{ps}\ll 1$. In this situation, $\nu \epsilon_{ps}$ of order
unity is enough to ensure that $\nu_{s} \gg 1$ and:
 \begin{equation}
\nu_{s}(c) = \nu \epsilon_{ps} (c-c^{\star})^{1/3} [c(1-c)] ^{1/3}
\epsilon_{p}^{-2/3}
\label{eq:scaling}
\end{equation}

In the strong coupling limit, the binodal line is also proportional to the
coupling parameter $\nu\epsilon_{ps}$. Numerical solution of
eqs.~(\ref{eq:spin0})-(\ref{eq:crit0}) indicates that the behavior of the
binodal line is simpler than the one of the spinodal line. In particular,
if the strong coupling limit can be defined only over a range of surfactant
coverage $c$ (for instance, when the special transition line intersects
the phase diagram), then the binodal line is found to be proportional to
$\nu\epsilon_{ps}$ for all values of $c$ (except for the limiting values
$c \rightarrow 0$ or for $c \rightarrow 1$). Figure 3 shows the dependence
of the binodal temperature on the special transition value $c^{\star}$, at
a fixed value of the polymer interaction parameter $\epsilon_{p}$. Lower
values of the special transition coverage $c^{\star}$ correspond, for a
fixed coupling $\nu\epsilon_{ps}$, to a higher binodal temperature.

\subsection{High polymer flexibility: $\epsilon_{p}\ll 1$; the possibility
of a triple point.}
 As was explained in Sec. 2, $\epsilon_{p}$ is likely to be very small for
a semi-dilute polymer solution. When $\epsilon_{p} \ll 1$, the strong
coupling limit is obtained for $c>c^{\star}$ (as was explained in Sec 3.5)
and the polymer is strongly adsorbed at the interface (see
Fig.~3). On the other hand,
for $c<c^{\star}$, the polymer is depleted from the interface and the
situation is a one of low coupling. Consequently, the features of the
phase diagram are particularly sensitive to the position of the special
transition line: when $c^{\star}>1$, the phase diagram is very close to
the one of the pure surfactant monolayer (see Fig.~1), while for
$c^{\star}<0$ the spinodal line is given by eq. (\ref{eq:scaling}) and
the binodal temperature is strongly shifted upwards.
When the special transition line intersects the phase diagram,
there is a competition between the two types of phase behavior. In Sec.
3.5, it was mentioned that the phase diagram of the strong coupling is
dominant in that situation.

More precisely, an expansion of the free energy in powers of
$\epsilon_{p}/\epsilon_{ps}$ shows that, for $c>c^{\star}$,
$F(c)\simeq-1/6(\epsilon_{ps}^{3}/\epsilon_{p}^{2}) (c-c^{\star})^{3}$,
while for $c<c^{\star}$, it yields $F(c)\simeq F_{s}(c)$. Fig.~4a shows
a plot of the free energy in a case where this sudden decrease of $F(c)$
for $c>c^{\star}$ is particularly well defined.
When $c^{\star}>1/2$, the binodal line of the pure surfactant monolayer
can be obtained with a common tangent construction (in the range of
temperatures where the two coexistence concentrations are lower than the
special transition coverage $c^{\star}$).
However, it corresponds to a metastable
state. There is yet a second pair of coexisting concentrations which
leads even to a lower free energy.

When $0<c^{\star}<1$, the phase diagram of the strong coupling limit
predominates over the one of the pure surfactant monolayer, while for
$c^{\star}>1$ this is not the case anymore. For physical values of the
parameters, the transition between the two regimes, occurring around
$c^{\star}=1$, is smooth. In some situations, the two pairs of coexisting
concentrations mentioned above are stable and
correspond to two first-order phase transitions, as is
shown in Fig.~4b. The total phase diagram exhibits two critical points and
one triple point. An example of such a
phase diagram with a triple point is
presented in Fig.~5. In Fig.~6 we show for the same set of parameters
three typical isotherms plotted in the reduced surface pressure $\Pi$ -- reduced
area per surfactant plane. The reduced surface pressure is defined by rescaling
the actual surface pressure ({\it i.e.},
the difference between the bare water/air surface tension
and that of the  surfactant monolayer) by $A_0\nu/k_B T$,
resulting in the relation
$\Pi =  \nu  c^2 \partial({F/c})/\partial{c}$. Depending on
the temperature, the isotherms can have zero, one or two plateaus corresponding
respectively to zero, one or two coexisting
regions.


\section{Discussion}
 In order to derive our model we assumed several simplifying assumptions.
The expression for the surfactant free energy $F_s$ does not take into
account the surfactant hydrophobic tails degrees of freedom. The coupling
between the tail conformations and the concentration of the surfactant
molecules becomes crucial at high surfactant densities and can lead to a
rich phase behavior \cite{Knobler,Kaganer}, which was not addressed here.
We consider here only the dilute phases of surfactant monolayers: the
gaseous phase at low densities and temperatures, the liquid-expanded phase
at higher densities and the condensation transition between them
\cite{Ben-Shaul}.

Another limitation of the model comes from our mean-field treatment of the
polymer free energy $F_{p}$. The assumption that the polymer solution is
semi-dilute may break down close to the interface if the polymer strongly
adsorbs. Our approach assumes that $c_{p}(z=0) \ll 1$ or equivalently
$\lambda \ll \rho$, where $\rho$ is the volume fraction of the monomers in
the bulk and $\lambda = \epsilon_{ps} (c-c^{\star})$. The coupling term
$F_{ps}$ should be regarded as the first term in an expansion. A more
precise study should take into account higher order terms, particularly in
the surfactant concentration $c$.

For water soluble polymers, hydrogen bonds play an important role because
of the strong polarity of water molecules and more refined expressions for
$F_p$ and $F_{ps}$ should be used. For example, the phase diagram of water
soluble polymers like PEO exhibits closed loops of immiscibility and the
definition of good solvent conditions is somewhat vague \cite{Napper}.

We are not aware of many experiments performed on adsorption of polymers
on a Langmuir monolayer, which will allow a direct comparison with our
results \cite{Benita}.  Interesting polymers that can be used
experimentally to test our theory are hydrophobically modified water
soluble polymers (HM-WSP) \cite{Iliopoulos},\cite{Wang}--\cite{Chari}.
These {\it associating}
copolymers are attracted by
the bare interface because of short hydrophobic side chains,
attached covalently to the main chain. By adjusting the number and the
length of the side chains, one can directly modify their surface affinity.
In our model it corresponds to $\gamma_{0}$, the parameter of interaction
between the polymer and the bare surface.

Figure 7 shows two different
kinds of interaction between the monomers and the interface. In (a), monomers
interact with the interface through any kind of short-ranged interaction:
either attractive or repulsive. In (b), the
attraction of HM-WSP polymers towards the interface is illustrated: one of
the aliphatic side chains of a HM-WSP polymer is in a low energy state in
the air subphase. The monomers of the main chain covalently bound to this
aliphatic group are consequently attracted in the region of the interface.

The
HM-WSP polymers can be used to study systematically the dependence of the
phase diagram on the special transition concentration $c^{\star}$ (as well
as on the coupling parameter $\epsilon_{ps}$), including the interesting
situation where the special transition line intersects the phase diagram
($0<c^{\star}<1$) because the polymer is repelled from the surfactant
molecules ($\alpha_{0}<0$). The surfactant can be chosen as nonionic, with
a polar head (hydrophilic) identical to the monomers, resulting in a
repulsion between the ``brush" (formed by the polar heads) and the polymer
(water being a good solvent for both) \cite{Safran}. The resulting
coupling parameter $\epsilon_{ps}$ is negative and the triple point occurs
for values of $c^{\star}$ close to zero and not close to unity as when
$\epsilon_{ps}>0$ (Fig.~5). However, such polymers usually have a
complicated behavior in the bulk and their self-assembly properties may be
crucial. Another possible candidate for experiments may be a polymer with
a ``simpler" behavior in water, like PEO, having a surface affinity
\cite{Bailey}. Probably here one needs to treat more explicitly the
hydrogen bonds.

The parameters in the model can also be tuned by changing the surfactant
characteristics. Pentadecanoic acid, for example,
whose gaseous to liquid
expanded transition has already been studied by several authors
\cite{Kim,Pallas}, can probably be an interesting surfactant to use.

In a previous work \cite{deGennes2}, several phase diagrams for the mixed
polymer-surfactant system have been proposed from qualitative and general
arguments. The proposed phase diagrams exhibit the special transition line
for the polymer ($c=c^{\star}$) and the coexistence line for the
surfactant molecules, but the interaction between these two lines was not
treated in detail. Our results suggest that the position of the special
transition line has a very strong effect on the position of the
coexistence line.  Ref. \cite{deGennes2} also predicted a {\it $\Theta$
line} separating a region where
the polymer segregates from the surfactant from a region where the
polymer and surfactant are mixed together.
Our mean-field model does not
address directly this prediction since we have only one minimum of the polymer
surface value $\phi_{s}$ as a function of surfactant coverage $c$ [see
eq.~(\ref{eq:phi})], and we assume that the polymer solution is
homogeneous in the directions parallel to the interface as long as the
surfactant monolayer is homogeneous.

>From our study it seems that the $\Theta$ line of
Ref. \cite{deGennes2} and the coexistence line are the same: regions of
different concentrations for the polymer correspond to regions of
different concentrations for the surfactant. The polymer can be attracted
to the interface and segregate from the surfactant only if it is attracted
by the bare interface but repelled by the surfactant molecules. This
situation is driven by energy terms which are first order in the polymer
concentration [see eq.~(\ref{eq:coupling})] and is consequently different
from the one predicted in Ref. \cite{deGennes2}. It will be interesting to
understand in a detailed way this discrepancy, especially for $0<c^\star
<1$, where the coupling between the special transition and the surfactant
phase diagram leads to the richest variety of phenomena.

\section{Conclusion}
 We addressed in detail the adsorption of a semi-dilute polymer solution
on a surfactant monolayer, and the resulting phase diagrams. In our model,
the most important degree of freedom is the local concentration $c$ of
surfactant at the interface. Since the monomers interact with
regions of different concentrations with an energy proportional to
$c-c^{\star}$ ($c^{\star}$ is the concentration at the {\it special
transition}), a rich variety of phenomena results from the coupling
between the polymer solution and the surfactant monolayer. The polymer
surface excess is enhanced and the phase diagrams of the surfactant
monolayer are modified.  The monomers induce an additional indirect
attraction between the surfactant molecules depending on the concentration
of surfactant on the interface. Consequently, the range of the homogeneous
region in the phase diagram decreases. When the monomers interact more
favorably with the surfactant molecules than with the bare interface, the
critical concentration itself increases.

The value of the special transition coverage $c^{\star}$, describing the
interaction of the monomers with the surfactant molecules relatively to
their interaction with the bare water/air interface, has a major effect on
the phase diagram of the surfactant monolayer. When the monomers are
repelled by the surfactant molecules as well as by the bare interface
(for instance, $\epsilon_{ps}>0$ and $c^{\star}>1$), the phase diagram is
not very different from the one of a simple surfactant monolayer without
polymer in the water subphase. On the other hand, when the monomers are
attracted by the surfactant molecules as well as by the bare interface
(for instance, $\epsilon_{ps}>0$ and $c^{\star}<0$), the increase of the
two-phase region can be important. In the intermediary situation both when
the polymer is attracted to the interface or depleted ($0<c^{\star}<1$),
these two scenarios are in competition leading to a rich phase behavior
and in some cases, the phase diagram displays two critical points and one
triple point (as in Fig.~5 and 9k).

Finally, we mention two possible extensions of the present study.
The first is to consider the situation of
soluble surfactants,
and to take into account the complex surfactant-polymer
interactions
in the bulk. Interesting experimental results have been obtained
for such systems\cite{Chari}.
Moreover, our model can easily be adapted to mixed monolayers with two
components. Such binary mixtures have been shown to exhibit immiscibility
at room temperature, ({\it e.g.}, for cholesterol and DMPC
\cite{Subramaniam,Seul}). The understanding of the features of this
phase transition may help to understand interactions
of polymers with flexible membranes (lipid bilayer)
and the phenomenon of {\it budding} of membranes
\cite{Sackmann1,Bloom,Sackmann2,Kawasaki}.

\bigskip
\noindent{\em Acknowledgments}
\medskip


We greatly benefited from discussions and correspondence
with P.-G. de Gennes, H. Diamant, J. F. Joanny,
B. Menes and S. Safran. Partial support
from the German-Israel Foundation (G.I.F) under grant No.
I-0197 and the US-Israel Binational Foundation (B.S.F.) under grant
No. 94-00291 is gratefully acknowledged. One of us (X.C.) acknowledges
support from the French Ministry of Foreign Affairs as well as from the
E.N.S. Lyon.

\pagebreak




\pagebreak
\setcounter{equation}{0}
\appendix
\section{Appendix A: The Spinodal Equation}
The general equation of the free energy of the system, as was derived in
Sec. 2, is
\begin{equation}
   F = \nu^{-1} c(1-c) +c\log{c} +(1-c)\log{(1-c)} -
\frac{\epsilon_{p}}{6} \OP \phi_{s}^{3} +3\phi_{s} -4 \CL
\label{eq:energA}
\end{equation}
 If we do not assume any specific temperature dependence for $\nu$,
$\epsilon_{p}$, $\epsilon_{ps}$ and $c^{\star}$, the global phase diagram
has to be studied in the five dimensional space of $\nu$, $\epsilon_{p}$,
$\epsilon_{ps}$, $c^{\star}$ and $c$. The binodal as well as the spinodal
surfaces are four-dimensional hypersurfaces in this five-dimensional space.
We shall consider both positive and negative $\nu$. It is worthwhile to
investigate the spinodal surface because it can be done analytically, and
it roughly describes the phase separation region, since the spinodal
surface lies always within the coexistence region. The surface of
instability (spinodal surface) also indicates the locations of possible
critical points. For fixed values of $\epsilon_{p}$, $\epsilon_{ps}$ and
$c^{\star}$, the spinodal equation (\ref{eq:spin0}) describes cuts of the
spinodal surface by the curve $\nu^{-1}=\nu^{-1}(c)$
\begin{equation}
   2\nu^{-1}=\frac{1}{c(1-c)} - \frac{\epsilon_{ps}^{2}}{\epsilon_{p}}
\frac{\phi_{s}^{3}}{\phi_{s}^{2}+1}
\label{eq:spiness}
 \end{equation}
 It is interesting to look also at other cuts of the parameter space
beside the ($\nu^{-1},c$) direction. An analytical expression can also be
found in the ($\epsilon_{p},c$) direction (but not for the other
interaction parameters).

Defining $u$ as
\begin{equation}
u= \frac{c-c^{\star}}{\epsilon_{ps}} \OP \frac{1}{c(1-c)} - 2\nu^{-1} \CL
\label{eq:Ap}
\end{equation}
 we note that $u=0$ for both the special transition line and the spinodal
line of the non-coupling case ($F(c)=F_{s}(c)$). From the spinodal
equation (\ref{eq:spiness}) it follows that, if $\epsilon_{p}$ is a
solution of the spinodal equation (\ref{eq:spin0}), $\epsilon_{p}^{2}$ is
the solution of a bi-quadratic equation:
\begin{equation}
   4u^{2}\epsilon_{p}^{4}- \lambda \OP 1+4u-u^{2} \CL
\epsilon_{p}^{2} - \lambda^{4} u =0
\label{eq:quadra}
\end{equation}
 Alternatively, it can be shown that, if $\epsilon_{p}$ is a solution of
(\ref{eq:quadra}), then, either $\epsilon_{p}$ or $-\epsilon_{p}$ is a
solution of the spinodal equation (\ref{eq:spin0}). As the polymer is
supposed to be in good solvent conditions, we disregard negative solutions
of the spinodal equation for $\epsilon_{p}$. The second order equation for
$\epsilon_{p}^{2}$ (\ref{eq:quadra}) may have either only one real
solution, or even no real solution at all, depending on its discriminant.
Furthermore, only positive solutions should be taken into account, since
$\epsilon_{p}^{2}>0$.  All these considerations give several bounds to the
possible values of $u$ and $\lambda$. The roots of the bi-quadratic
equation (\ref{eq:quadra}) are:
 \begin{equation}
   \beta_{\pm}=\frac{\lambda^{2}}{8u^{2}} \OP 1+4u-u^{2}
\pm (u+1)\sqrt{1+6u+u^{2}} \;\CL
\label{eq:roots}
\end{equation}
 When $\beta_{+}$ is real and positive, only one solution of the spinodal
equation exists: either $\epsilon_{p}=\sqrt{\beta_{+}}\;$ or
$\;\epsilon_{p}= -\sqrt{\beta_{+}}$. We shall consider this solution only
in the case when it is positive. The same can be said for $\beta_{-}$.

Defining $u_{0}=-3+\sqrt{8} \simeq -0.1716$, we separate three cases for the
roots of (\ref{eq:quadra}) given by eq.~(\ref{eq:roots}):
\begin{itemize}
\item For $u>0$, $\beta_{+}$ is real and positive (while $\beta_{-}$ is
negative, and therefore,
irrelevant). Hence, there is only one solution for $\epsilon_{p}$.
\item For $u_{0}\leq u \leq 0$, $\beta_{+}$ and $\beta_{-}$ are both real
and positive
(for $u \rightarrow 0$, $\beta_{-}  \rightarrow 0$). Hence,
there are  two solutions for
$\epsilon_{p}$.
\item For $u<u_{0}$, neither
$\beta_{+}$ nor $\beta_{-}$ are real and positive. Hence, no physical
solution for $\epsilon_p$.
\end{itemize}

Once $\epsilon_{ps}$ and $c^{\star}$ are fixed, $\epsilon_{p}$ is a
function of $\nu$ and $c$. It is instructive to identify the domains in
the ($\nu,c$) plane where the spinodal equation (\ref{eq:spin0}) has zero,
one or two real solutions. This requires to locate the lines $u=0$ and
$u=u_{0}$. Depending on the values of $\epsilon_{ps}$ and $c^{\star}$, two
different situations are identified and shown in Fig. 8. In particular, it
can be seen in Fig.~8b that, for $1/2<c^{\star}<1$, there can be a range
of values for $\nu^{-1}$ where cuts of the spinodal surface in the plane
($\epsilon_{p},c$) are disconnected (if $\epsilon_{ps}$ is below a certain
critical value). For instance, in Fig. 9j ($1/2<c^{\star}=0.86<1$), a cut
of the phase diagram at $\nu_0^{-1}<\nu^{-1} =1.8<\nu_1^{-1}$ is presented.
The spinodal line is defined for two disconnected regions of concentration:
one centered around $c=0.6$ and one at high concentrations.

\section{Appendix B: The Global Phase Diagram}

The phase diagram for the surfactant monolayer, including the surfaces of
instability and coexistence, is presented in the three dimensional
parameter space of $\nu^{-1}$, $\epsilon_{p}$ and $c$ through several cuts
in two dimensions ($\epsilon_{p}$ and $c$), while fixing the other two
parameters, $\epsilon_{ps}$ and $c^{\star}$. We first discuss some general
features of the phase diagram, starting with the simple case of no {\it
effective} coupling between the polymer and the interface
($\epsilon_{ps}=0$; $F(c)=F_{s}(c)$) as a reference point. The dependence
of the phase diagram on the position of the special transition line is
then studied, and the phase diagrams are presented for the cases:
$c^{\star}<0$, $c^{\star}>1$ and $0<c^{\star}<1$.

When there is no {\it effective} coupling between the polymer solution and
the surfactant monolayer, $\epsilon_{ps}=0$ and the special transition
line is not defined. The concentration of the polymer at the interface is
identical to its bulk value and depends on the strength of the interaction
between the monomers and the interface $\alpha_{0}=\gamma_{0}=\lambda$.
The phase separation in the monolayer was explained in Sec. 2.1, and all
its features are independent on the polymer parameter $\epsilon_{p}$. In
the plane ($\nu^{-1},c$), the line $\nu^{-1}=1/[2c(1-c)]$ delimits the
unstable region (see Fig.~8). From eq.~(\ref{eq:spiness}) it appears that
as soon as there is a non zero effective coupling, the line of instability
occurs for smaller values of $\nu^{-1}$. It means that the coupling
enlarges the region of instability of the homogeneous state. Consequently,
the dark shaded region inside the line $\nu^{-1}=1/[2c(1-c)]$ in Fig.~8 is
always a zone of instability, for any value of the parameters
$\epsilon_{p}$, $\epsilon_{ps}$ and $c^{\star}$. When $\epsilon_{p}$ is
big, the polymer ``stiffness" induces the polymers in solution to be
uniform, and the coupling acts as a chemical potential having a value
$\frac{1}{2}\epsilon_{ps}$, as can be seen from eq.~(\ref{eq:energy})
(after expanding $\phi_{s}$ to first order in powers of
$\epsilon_{p}^{-1}$). In this limit, the properties of the phase
separation are the same as in the non-coupling case. Considering the cuts
of the phase diagram in the direction ($\epsilon_{p},c$) for
$\epsilon_{ps}=0$, there are two different regions of the parameter space:
\begin{itemize}
\item the region $\nu^{-1}<2$, where there is no instability and no phase
separation.
\item the region $\nu^{-1}>2$, where there is a phase separation.
\end{itemize}
Fig.~9a shows the spinodal and the binodal lines for $\nu = 2.31$. Both lines
are vertical because the properties of the phase transition are independent
of $\epsilon_{p}$, as was already explained above.
In the limit $\nu^{-1}\rightarrow 2$, the phase diagram reduces to a line
of critical points parallel to the $\epsilon_{p}$ axis at $c=0.5$. The
coupling results in a deformation of the line of critical points and
enlarges the phase coexistence region. The cuts of the phase diagram in
the direction ($\epsilon_{p},c$) for $\nu^{-1}<2$ are topologically very
different from the cuts obtained for $\nu^{-1}>2$, because the latter
necessarily contain the region of instability of the non-coupling situation
(even for large values of $\epsilon_{p}$), while, for the former, large
values of $\epsilon_{p}$ necessarily correspond to domains of stability of
the monolayer.

We first discuss the cases where the special transition line $c=c^{\star}$
does not intersect the parameter space ($c^{\star}<0$ or $c^{\star}>1$).
The phase diagrams are constructed by numerical solution of
eqs.~(\ref{eq:spin0})-(\ref{eq:crit0}).

\subsection{The $c^{\star}<0$ case}
 When both the surfactant and the interface attract the monomers, two
situations can occur: if $\alpha_{0}>\gamma_{0}>0$, then $\epsilon_{ps}>0$
and $c^{\star}\leq 0$. But if $\gamma_{0}>\alpha_{0}>0$, $\epsilon_{ps}<0$
and $c^{\star} \geq 1$. However, as was explained at the beginning of Sec.
2.4, these two situations can be mapped onto each other. Therefore, we
concentrate here on the case $\epsilon_{ps}>0$ and $c^{\star}\leq 0$. The
only relevant solution of the spinodal equation for $\epsilon_{p}$ is then
$\epsilon_{p}=+\sqrt{\beta_{+}}$, defined outside of the region of
instability of the non-coupling case. When $\epsilon_{p}$ is big, the
polymer stiffness enforces the polymer solution to be uniform, $\phi_{s}
\simeq 1$, and the monomers exert a strong uniform attraction on the
surfactant molecules.

For $\nu^{-1}<2$, there is a phase separation, provided that
$\epsilon_{p}$ is small enough.  This is shown in Fig.~9b, where it can
be seen that the spinodal and the binodal lines join at a critical point.
When $\nu^{-1}$ is changed, the basic
topology of the binodal and spinodal lines remains unchanged. But, in the
limit $\nu^{-1}\rightarrow 2^{-}$, close to the region where a phase
separation occurs even without a coupling, the value of $\epsilon_{p}$ at
the critical point increases. Moreover, the critical concentration tends
to the value of the non-coupling case $c=1/2$ (because $\epsilon_{p}$ at
the critical point is large). If, on the other hand $\nu^{-1}$ decreases,
the value of $\epsilon_{p}$ at the critical point decreases and the
critical concentration $c_{c}$ increases, $c_{c} \rightarrow 1$ when
$\nu^{-1}\rightarrow -\infty$.

For $\nu^{-1}>2$, the phase diagram is presented in Fig.~9c. There is no
critical point because the phase separation occurs for all values of
$\epsilon_{p}$. For
$\epsilon_{p} \rightarrow 0$, the effects of the coupling between the
monomers and the surfactant molecules are the strongest: the polymer is
strongly attracted to the interface, particularly in the regions dense in
surfactant.  As a result, the surfactant molecules aggregate with a
close-packing coverage $c=1$.

\subsection{The $c^{\star}>1$ case}
 The opposite situation happens when both the surfactant and the bare
interface repel the monomers. As explained above it is enough to consider
only $\epsilon_{ps}>0$ and $c^{\star} \geq 1$. In this situation, the
function $\nu^{-1}(c)$ defined by the line $u=u_{0}$ in the plane
$(\nu^{-1},c)$ (see Appendix A) has a minimum $\nu^{-1}_{o}<2$, which is a
function of $\epsilon_{ps}$ and $c^{\star}$. When $\nu^{-1}<\nu^{-1}_{o}$,
then $u<u_{0}$. The spinodal equation has no solution and the homogeneous
monolayer state is stable. For $\nu^{-1}>\nu^{-1}_{o}$, from a topological
point of view, the spinodal surface is deformed for intermediate values of
$\epsilon_{p}$. Whereas for large values of $\epsilon_{p}$, the polymer
remains stiff and the properties of the phase transition are those of the
non-coupling case, for small values of $\epsilon_{p}$, the repulsion of the
interface is dominant. The polymer is strongly depleted and the coupling
has no effect.

The cuts of the phase diagram in the direction ($\epsilon_{p},c$) have the
following features:
 \begin{itemize}
\item For $\nu^{-1}_{o}<\nu^{-1}<2$ the phase diagram has a closed loop
immiscibility curve with an
upper and a lower critical points as shown on Fig.~9d. When
$\nu^{-1}\rightarrow 2^{-}$, the
upper critical point tends to $\infty$ and the lower one to $0$, whereas when
$\nu^{-1}\rightarrow \nu^{-1}_{o}$, the domain of instability becomes very
small.
\item For $\nu^{-1}>2$, there is a phase separation, whose
characteristics
differ from the one of the non-coupling case only for intermediate values of
$\epsilon_{p}$. An example of such a case is presented in Fig.~9e.
\end{itemize}

The line of critical points has only, for intermediate values of
$\epsilon_{p}$, a small distortion with respect to the straight line of
the non-coupling case. One consequence is that there is a maximal value
for the critical concentration. When $c^{\star}$ is increased, the
repulsion from the interface increases and the phase diagram resembles
the non-coupling case, even at intermediate values of
$\epsilon_{p}$.

\subsection{The $0<c^{\star}\leq 1$ case}
 We turn now to the more complex case where the special transition line
intersects
the physical range of parameter space
($0<c<1$). The situation for $0<c^{\star}<1/2$ is
rather simple, and two typical phase diagrams are shown in Figs.~9f (for
$\epsilon_{ps}<2$) and 9g and (for $\epsilon_{ps}>2$). They resemble the
$c^{\star}<0$ case. However, for $1/2<c^{\star}<1$, depending on the value of
the coupling parameter $\epsilon_{ps}$, the competition between the
different terms in the free energy leads to a large variety of possible
phase diagrams. An example of a phase diagram with a reentrant phase is
presented in Fig.~9h. In Fig.~9i yet another
type of a phase diagram is presented where
a second metastable region resides
inside the two phase region.
In some cases, as was
suggested at the end of Appendix A, the phase diagram appears as if
it is composed of two disconnected parts (those are cuts
obtained for $\nu_0^{-1}<\nu^{-1}<2$). An example is given in
Fig.~9j. Note that for $2<\nu^{-1}<\nu_1^{-1}$ the spinodal surface is
disconnected while the binodal surface is connected, since at
very low values of $\epsilon_p$, a phase separation always occurs
between very dense ($c \simeq 1$) and very dilute ($c \simeq 0$) regions
(for $c^{\star}<1$), as is explained below.
In the three dimensions ($\epsilon_{s},\epsilon_{p},c$), the phase diagram is
always connected: when $\nu^{-1}$ is increased the disconnected parts of
the phase diagram meet and this may result in the appearance of a triple
point for a range of values for $\epsilon_{s}$ (an example is shown in
Fig.~9k). These triple points occur only for a small range of values for
$\epsilon_{p}$; typically, for $\epsilon_{p} \simeq 0.1$.

Note that in all these phase diagrams (Figs.~9f to 9k), the binodal line
at low values of $\epsilon_{p}$ always ends at $c=0$ and $c=1$. It can be
understood since for low values of $\epsilon_{p}$, regions of surfactant
concentrations $c<c^{\star}$ (and outside of the region of instability of
the non-coupling case) are not unstable because they repel the monomers,
and the coupling consequently does not have any effect. The free energy is
unchanged as compared to the non-coupling case. However, it is strongly
modified for $c>c^{\star}$, as has been explained in Sec. 3.6. and is
shown on Fig.~4a. The common tangent construction results in a couple
of stable coexistence points ($c\simeq 0$, $c\simeq 1$). Therefore, the
regions of surfactant concentrations $c<c^{\star}$ (and outside of the
region of instability of the non-coupling case) are metastable.

\setcounter{equation}{0}
\section{Appendix C: The Critical Point}

It is possible to obtain analytical results for the critical point, by
solving simultaneously eqs.~(\ref{eq:spin0}) and (\ref{eq:crit0}), for the
two unknowns: the critical concentration $c_c$
and the critical temperature.

\subsection{Bounds on the critical concentration}
 In general, no analytical expression for the critical concentration $c_c$
exists but we derive an analytical expression for an upper bound.
Equation~(\ref{eq:crit0}) can be rewritten as a third order equation for
$\phi_{s}^{2}$:
\begin{equation}
   \phi_{s}^{6} + 3 \phi_{s}^{4} + 3 f(c) \phi_{s}^{2} + f(c) =0
\label{eq:cefini}
\end{equation}
where the coefficient $f(c)$ is defined as:
\begin{equation}
  f(c)=\frac{1-2c}{(1-2c) + c^{2} (1-c)^{2}\epsilon_{ps}^{3}/\epsilon_{p}^{2}}
\label{eq:f(c)}
\end{equation}
 We denote $c_{1}$ as the value of the surfactant concentration for which
$f(c)$ diverges. A careful study of eq.~(\ref{eq:cefini}) shows that
$f(c)<0$, and yields a non-analytical solution for $\phi_{s}$ and
consequently for $c_c$. It can also be shown that $1/2<c_c<c_{1}$. From
its definition, $c_{1}$ is the solution of a fourth order equation and has
a (complicated) analytical expression as an odd increasing function of
$\epsilon_{ps}^{3}/\epsilon_{p}^{2}$. For
$\epsilon_{ps}^{3}/\epsilon_{p}^{2}=0$, $c_{1}=1/2$ and when
$\epsilon_{ps}^{3}/\epsilon_{p}^{2}\rightarrow\infty$, $c_{1} \rightarrow
1$ \cite{c1}.

\subsection{The critical point in the strong coupling limit}
 Hereafter we concentrate on the specific case of the strong coupling
limit, where $\nu^{-1}$ can be neglected in the spinodal equation
eq.~(\ref{eq:spin0})
 \begin{equation}
   \phi_{s}^{2}+1= \epsilon_{ps}^{2} \phi_{s}^{3}\,\,
\frac{c(1-c)}{\epsilon_{p}}
\label{eq:spin2}
\end{equation}
 Taking the polymer order parameter $\epsilon_{p}$ and the special
transition $c^{\star}$ as known parameters, the characteristics of the
critical point can be determined by solving a system of three equations
(the spinodal equation (\ref{eq:spin2}), eq.~(\ref{eq:crit0}) and the
definition of $\phi_{s}$ (\ref{eq:phi})) with three unknowns: the
concentration of surfactant $c$, the temperature appearing through
$\epsilon_{ps}$ and $\phi_{s}$ (which has been added for mathematical
convenience). First, bounds on the critical concentration are obtained,
depending only on the special transition concentration. Then, a method of
determination of the critical concentration is discussed. Using
eq.~(\ref{eq:spin2}), eq.~(\ref{eq:crit0}) can be rewritten as a second
order equation for $\phi_{s}^{2}$.

%
As there is at least one real and positive solution to
this equation, a bound on the value of the critical concentration,
depending only on the special transition concentration $c^{\star}$ is
obtained (Fig.~10).
%
\begin{itemize}
\item
For $c^{\star}\leq 0$: $\;\; 1- c^{\star} -\sqrt{c^{\star 2}-
c^{\star}+1} \leq c_c \leq \frac{1}{3} \left[
(c^{\star}+1)+\sqrt{c^{\star 2}-c^{\star}+1}
\right]$
\item
For $0\leq c^{\star}\leq 1$: $\;\;\frac{1}{3} \left[
(c^{\star}+1)-\sqrt{c^{\star 2}-c^{\star}+1} \right] \leq c_c \leq
\frac{1}{3} \left[ (c^{\star}+1)+\sqrt{c^{\star 2}-c^{\star}+1} \right]$
\item
For $c^{\star}>1$: $\;\;\frac{1}{3} \left[
(c^{\star}+1)-\sqrt{c^{\star 2}-c^{\star}+1} \right] \leq c_c \leq
1-c^{\star}+\sqrt{c^{\star 2}-c^{\star}+1}$
\end{itemize}
 For $c^{\star}=0$, this bound is $0<c_c<2/3$. Therefore, if the effective
coupling parameter $\epsilon_{ps}$ is positive, we know that the critical
concentration obeys $1/2<c_c<2/3$.

The second order equation for $\phi_{s}^{2}$ can be rewritten as a second
order equation for the critical concentration $c_c$, depending on the
special transition concentration $c^{\star}$ and on the concentration of
monomers at the interface $\phi_{s}$:
 \begin{equation}
[4-3 (\phi_{s}^{2}+1)^{2}] c^{2} - 2 \left[ 2 -
(1+c^{\star})(\phi_{s}^{2}+1)^{2} \right] c -
c^{\star} (\phi_{s}^{2}+1)^{2}=0
\label{eq:crit2}
\end{equation}
The study of this equation shows that:
\begin{itemize}
\item For $c^{\star}=0$ or $c^{\star}=1$, it reduces to a first order
equation.
\item
For $c^{\star}<0$,
$\;\; c_c= c_{-} \equiv$

$ \left( 2-(1+c^{\star})(\phi_{s}^{2}+1)^{2} -\sqrt{(\phi_{s}^{2}+1)^{4}
(c^{\star 2}-c^{\star}+1) -4(\phi_{s}^{2}+1)^{2}+4}\,\,\right)/\OP
4-3(\phi_{s}^{2}+1)^{2}\CL$
\item
For $c^{\star}>1$,
$\;\; c_c= c_{+} \equiv $

$\left( 2-(1+c^{\star})(\phi_{s}^{2}+1)^{2} +\sqrt{(\phi_{s}^{2}+1)^{4}
(c^{\star 2}-c^{\star}+1) -4(\phi_{s}^{2}+1)^{2}+4}\,\,\right)/\OP
4-3(\phi_{s}^{2}+1)^{2}\CL$
\item
 For $0<c^{\star}<1$, the situation is more complicated. It can be shown,
in particular, that there is a minimum for the concentration of the
monomers at the interface at the critical value. This minimal value always
corresponds to a situation of depletion for the polymer solution. It can
also be shown that in some situations, when $\epsilon_{ps}>0$ and
$1/2<c^{\star}<1$ or symmetrically when $\epsilon_{ps}<0$ and
$0<c^{\star}<1/2$, the solutions $c_{-}$ and $c_{+}$ may be relevant: this
accounts for the possibility of two critical points.
\end{itemize}

The definition of $\phi_{s}$ (\ref{eq:phi}) can be rewritten as:
\begin{equation}
   \phi_{s}^{2}-1=\frac{\epsilon_{ps}}{\epsilon_{p}} \phi_{s} (c-c^{\star})
\label{eq:phi2}
\end{equation}
 $\epsilon_{ps}$ can be eliminated from the spinodal equation
(\ref{eq:spin2}) by using its expression obtained from
eq.~(\ref{eq:phi2}). On the other hand, the concentration can be
substituted by using its expression $c_{-}(\phi_s ,c^{\star})$ or
$c_{+}(\phi_s ,c^{\star})$. One is left with a high order polynomial
equation for $\phi_{s}$ depending only on the special transition
concentration $c^{\star}$ and the polymer interaction parameter
$\epsilon_{p}$. Once this equation is numerically solved, the other
characteristics of the critical point (concentration, temperature) are
easily obtained as simple functions of $\phi_{s}$. In particular, once
the critical concentration is known, the critical $\epsilon_{ps}$,
hence the critical temperature are obtained from eq. (C.5).

\pagebreak
\pagestyle{empty}
\centerline{\bf Figure Captions}
\bigskip
\begin{itemize}

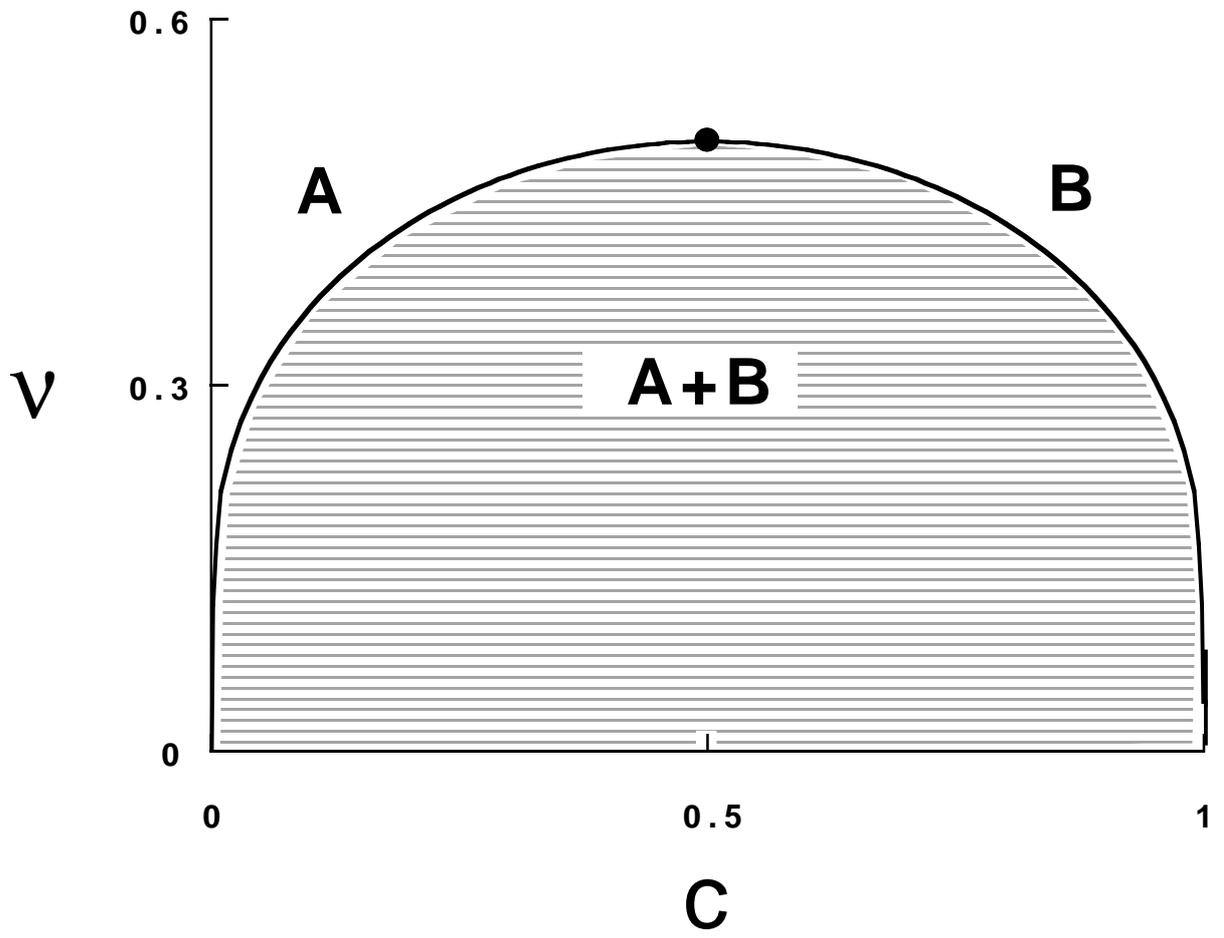
 \item{\bf Figure 1:} The phase diagram for a bare surfactant monolayer,
without polymer in the water subphase. At low $\nu$ (low temperatures),
the homogeneous state is unstable and the binodal line delimits the
two-phase coexistence region labeled A+B. The critical point is located at
$\nu_c=0.5$, $c_c=0.5$ and is shown by a dot.

\item{\bf Figure 2:}
 (a) The polymer order parameter $\phi_{s}$ and (b) the polymer surface
excess, $\tilde{\Gamma} \equiv\Gamma/(c_{b}\xi)= \phi_{s}-1$, as function
of the surfactant coverage $c$ for $\epsilon_{ps}/\epsilon_{p} = 10$ and
$c^{\star}=0.5$. In (a) the special transition line divides the region
where the polymer is adsorbed ($c>c^{\star}$) from the region where it is
depleted ($c<c^{\star}$).
 In (b), for homogeneous monolayers of concentrations $c_{1}=0.29$,
$c_{2}=0.875$ and $c^{\star}=0.5$, the polymer surface excess is
respectively $-0.6$, $3$ and $0$. Due to the convexity of the curve, for a
surfactant monolayer of average concentration $c^{\star}$ demixing between
coexisting regions of concentration $c_{1}$ and $c_{2}$, the surface
excess $\Gamma$ is positive.

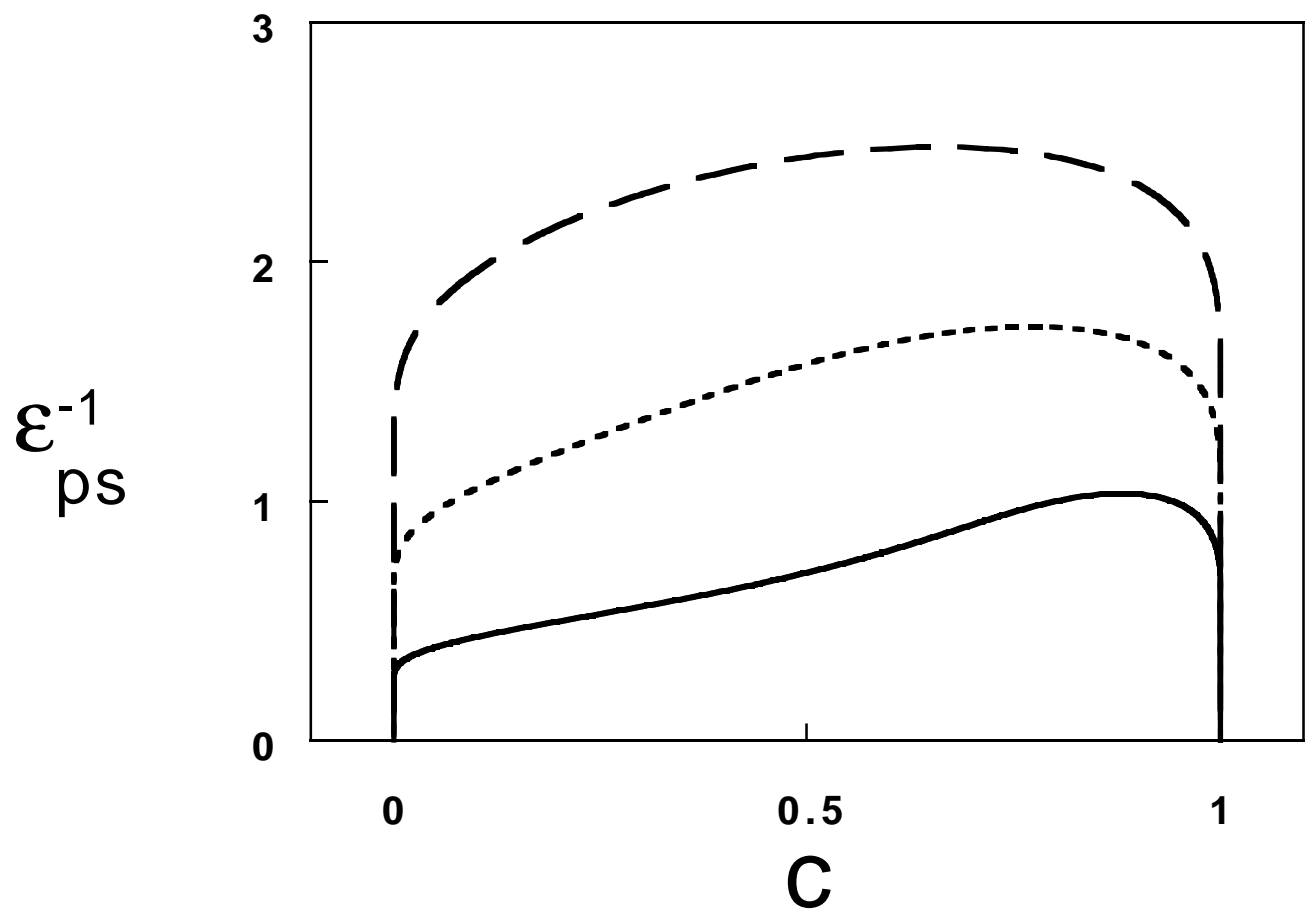
\item{\bf Figure 3:} The value of $\epsilon_{ps}^{-1}$ on the binodal line
in the strong coupling limit (independent on $\nu$) for $c^{\star}=0$
(dashed line), $c^{\star}=0.5$ (dotted line) and $c^{\star}=0.8$ (full
line) at a fixed value of the polymer interaction parameter
$\epsilon_{p}=0.1$. The binodal temperature $\nu$ is directly related to
$\epsilon_{ps}^{-1}$ by $\nu=(\nu \epsilon_{ps})\epsilon_{ps}^{-1}$. When
$c^{\star}$ decreases, the polymer interacts more favorably with the
interface and its effect on the phase diagram of the surfactant monolayer
is stronger: the binodal temperature increases.

\item{\bf Figure 4 :} The free energy $F$ as a function of the surfactant
concentration $c$ at the interface for $\nu=0.48$, $\nu\epsilon_{ps}=1$,
$c^{\star}=0.95$. In (a) $\epsilon_{p}=0.02$ and in (b)
$\epsilon_{p}=0.1$. The common tangent construction is shown by thin lines
and the coexistence values by circles. In (a), in the region $0<c<c^{\star}$,
the plot is almost similar to the case of the pure surfactant monolayer. The
corresponding
coexistence region, $0.33 \leq c \leq 0.67$ is only metastable, since it
is contained in the second
coexistence region $0.11 \leq c \leq 1.0$. In (b), both coexistence
regions $0.31 \leq c \leq 0.72$ and $0.83 \leq c \leq 0.99$ occur.

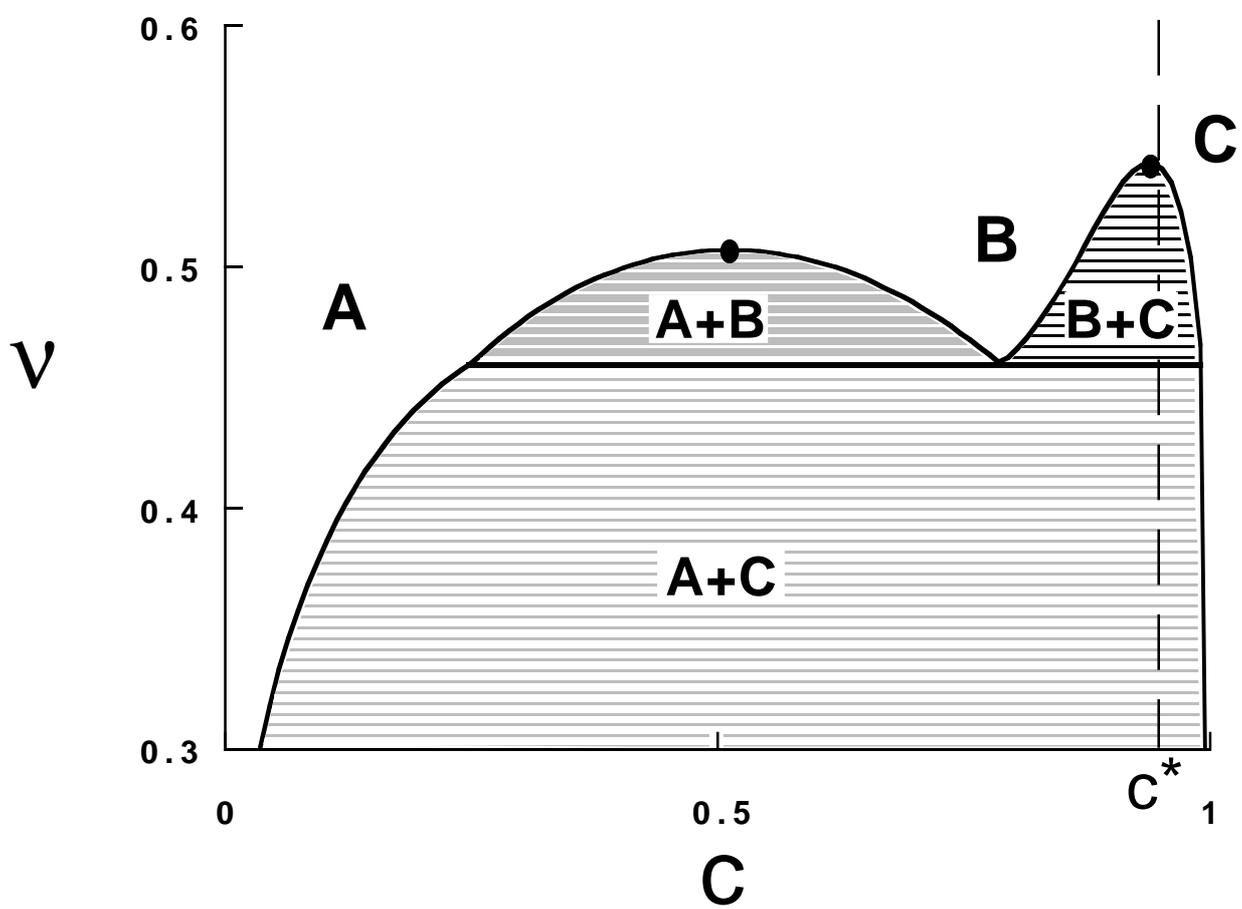
\item{\bf Figure 5:} The phase diagram of the surfactant monolayer for
$c^{\star}=0.95$, $\epsilon_{p}=0.1$ and $\nu \epsilon_{ps}=1$. The
two-phase region labeled A+B ends at the critical point: $\nu_c=0.51$,
$c_c=0.50$. The second B+C critical point is located at $\nu_c=0.53$,
$c_c=0.94$. All three two-phase regions: A+B, A+C and B+C join at a {\it
triple point}: $\nu=0.46$, $c_A=0.25$, $c_B=0.79$ and $c_C=0.99$, where
all three phases (A, B and C) coexist. Critical points are shown by a dot.

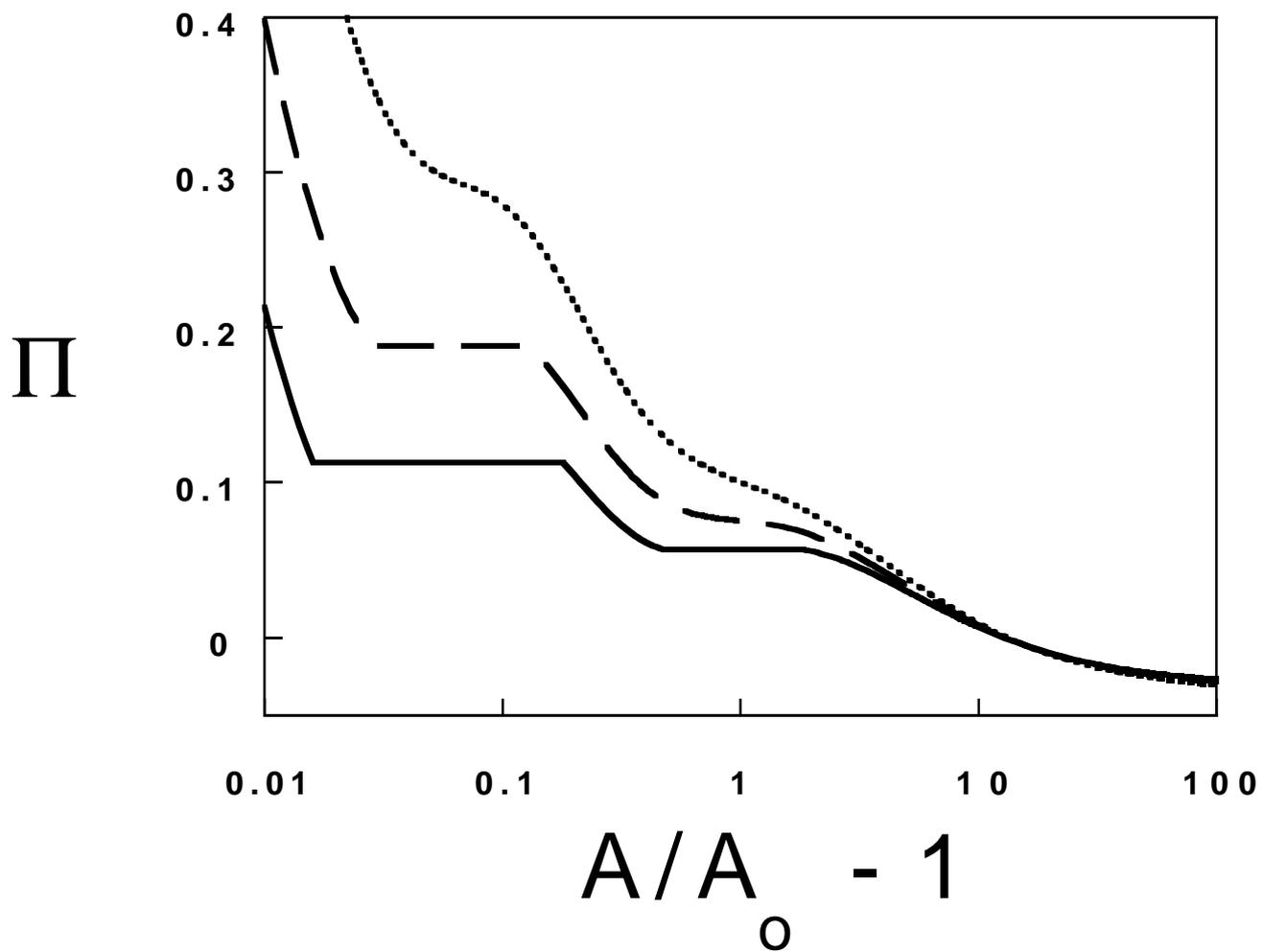
\item{\bf Figure 6:} Isotherms for the surfactant monolayer. The reduced
surface pressure $\Pi =  \nu c^2
\partial{(F/c)}/\partial{c}$ is plotted versus
the reduced
area per molecule $1/c - 1=A/A_0 -1$ on a logarithmic scale.
$A_0$ is the close-packing area and $F$ is the total free energy as
defined in the text. Three typical
isotherms are shown for three different temperatures:
$\nu =0.56$ (dotted line, no phase transition),
$\nu=0.52$ (dashed line, one phase transition), $\nu=0.49$ (full line, two
phase transitions). The other parameters are identical to the ones of
Fig. 5.

\item{\bf Figure 7:}
Two different
kinds of interaction between the monomers and the interface. In (a),
monomers interact with surfactant molecules either attractively
(for instance, as a
consequence of favorable van der Waals interactions) or repulsively
(if, on the contrary, water is a good solvent for both molecules, and the
surfactant heads act as a polymer brush). In (b), an associating polymer
is shown; the hydrophobic sub-chains tend to dispose themselves in the air
subphase and consequently attract the polymer chain close to the
interface.
(adapted from Ref. \cite{deGennes2})

\item{\bf Figure 8:} Regions of definition of the spinodal hypersurface in
the plane $(-\nu^{-1}, c)$ for:  (a) $\epsilon_{ps}=1$ and $c^{\star}=0.8$;
(b) $\epsilon_{ps}=1$ and $c^{\star}=0.95$. The full line is the $u=u_{0}$
line defined in Appendix A. The curved dashed line is the line of
instability for the non-coupling case: $2\nu^{-1}= 1/[c(1-c)]$; the
vertical dashed line is the line $c=c^{\star}$. The region where only the
solution $\beta_{+}$ is real and positive, and generates a positive solution
$\epsilon_{p}$ to the spinodal equation is light shaded, while the region
where both solutions $\beta_{+}$ and $\beta_{-}$ are real and positive,
and
generate a positive solution $\epsilon_{p}$, is hashed. The dark shaded
region is the zone where the surfactant monolayer is always unstable. The
region above the full line ($u=u_0$)
is a zone where neither $\beta_{+}$ nor $\beta_{-}$ are real
and positive. In (b), the $u=u_{0}$ line has two extrema at $\nu=\nu_{o}$
and $\nu=\nu_{1}$.

\item{\bf Figure 9:} Cuts of the phase diagram in the $(\epsilon_{p},c)$
plane. The full lines represent the binodal surface, while the dashed
lines represent the spinodal surface. Critical points are shown by a full
circle. (a) $\nu^{-1}=2.31$ and $\epsilon_{ps}=0$; this plot shows the
reference situation of the non-coupling case. Next plots illustrate the
text of Appendix B, in the three cases $c^{\star} \leq 0$, $c^{\star} \geq
1$ and $0 \leq c^{\star} \leq 1$. In (b) to (j) $\epsilon_{ps}=1$. (b)
$\nu^{-1}=1$ and $c^{\star}=-1$; (c) $\nu^{-1}=2.2$ and $c^{\star}=-1$;
(d) $\nu^{-1}=1.95$ and $c^{\star}=2$. This plot shows a closed loop phase
diagram. (e) $\nu^{-1}=2.2$ and $c^{\star}=2$. The $\epsilon_{p}$ scale in
this plot is logarithmic for convenience. (f) $\nu^{-1}=1$ and
$c^{\star}=0.3$; (g) $\nu^{-1}=2.2$ and $c^{\star}=0.3$; (h)
$\nu^{-1}=1.8$ and $c^{\star}=0.84$. This phase diagram displays a
reentrant phase. (i) $\nu^{-1}=2.1$ and $c^{\star}=0.84$. The
$\epsilon_{p}$ scale in this plot is logarithmic for convenience and the
shaded region corresponds to the metastable states. (j) $\nu^{-1}=1.8$ and
$c^{\star}=0.86$. This plot shows a disconnected phase diagram. In (k)
$\nu^{-1}=2.1$, $\epsilon_{ps}=2.1$ and $c^{\star}=0.95$. A first
two-phase region A+B extends to infinite values for $\epsilon_{p}$; the
second two-phase region ends at a critical point $c_c=0.90$,
$(\epsilon_{p})_c=0.16$; all three two-phase regions: A+B, A+C and B+C
join at a {\it triple point}: $\epsilon_{p}=0.06$, $c_A=0.30$, $c_B=0.70$
and $c_C=0.99$, where all three phases (A, B and C) coexist.

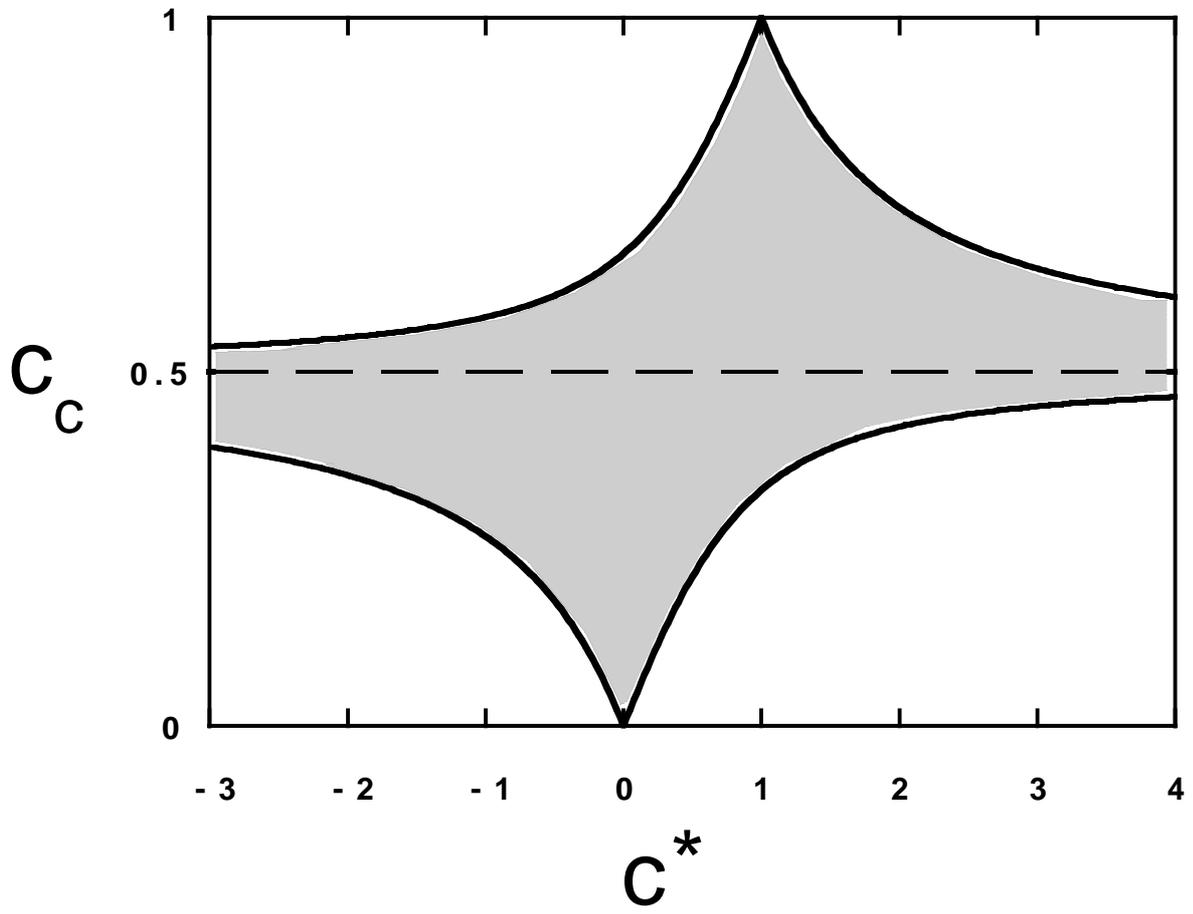
\item{\bf Figure 10:} In the strong coupling limit, the range of the
critical concentration $c_c$ depends only on the special transition
concentration $c^{\star}$, and is shown inside the shaded region.

\end{itemize}

\begin{thebibliography}{99}
\bibitem{smalleps} A more precise criterion for the validity of
eq.~(\ref{eq:lowexp}) is
that $\nu\epsilon_{ps}\ll \epsilon_{p}/(|c-c^{\star}|)$ and $\nu\epsilon_{ps}\ll
\epsilon_{p}^{1/2}$. For a semi-dilute polymer solution, $\epsilon_{p}$ is
in general very
small with respect to unity (as was mentioned in Sec. 2), and the low
coupling limit may
hold only for very small values of $\nu \epsilon_{ps}$.
\end{thebibliography}
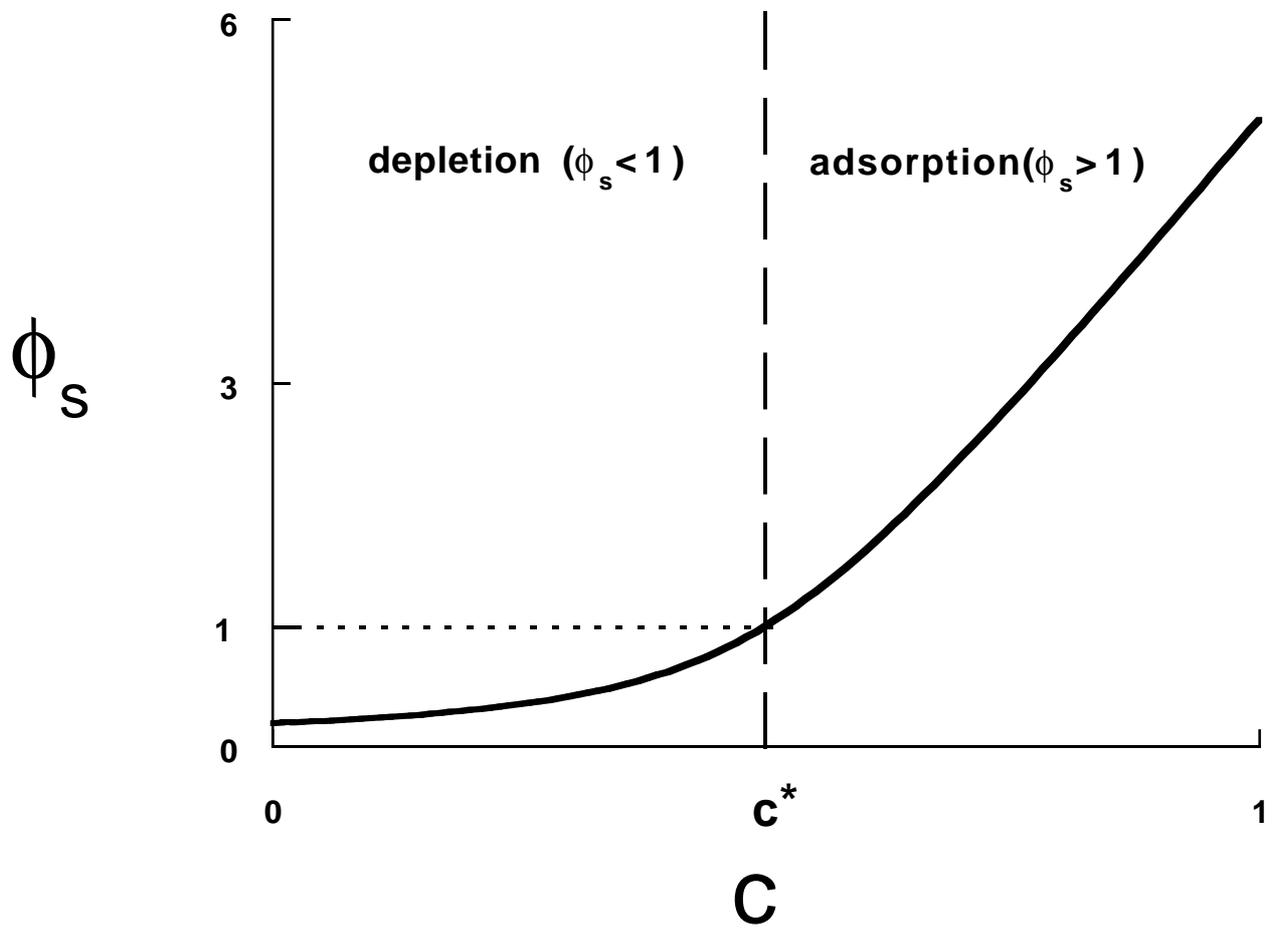
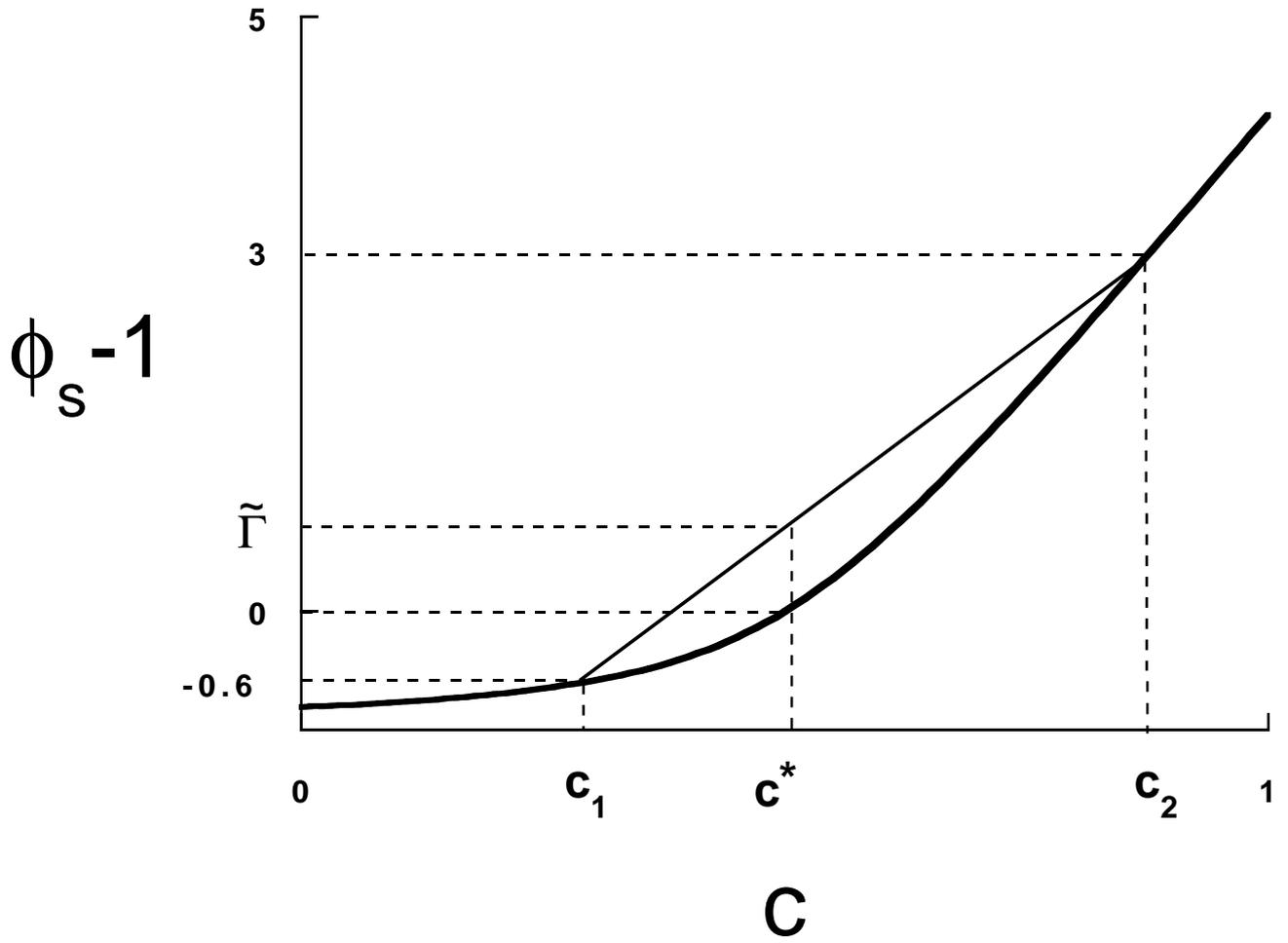
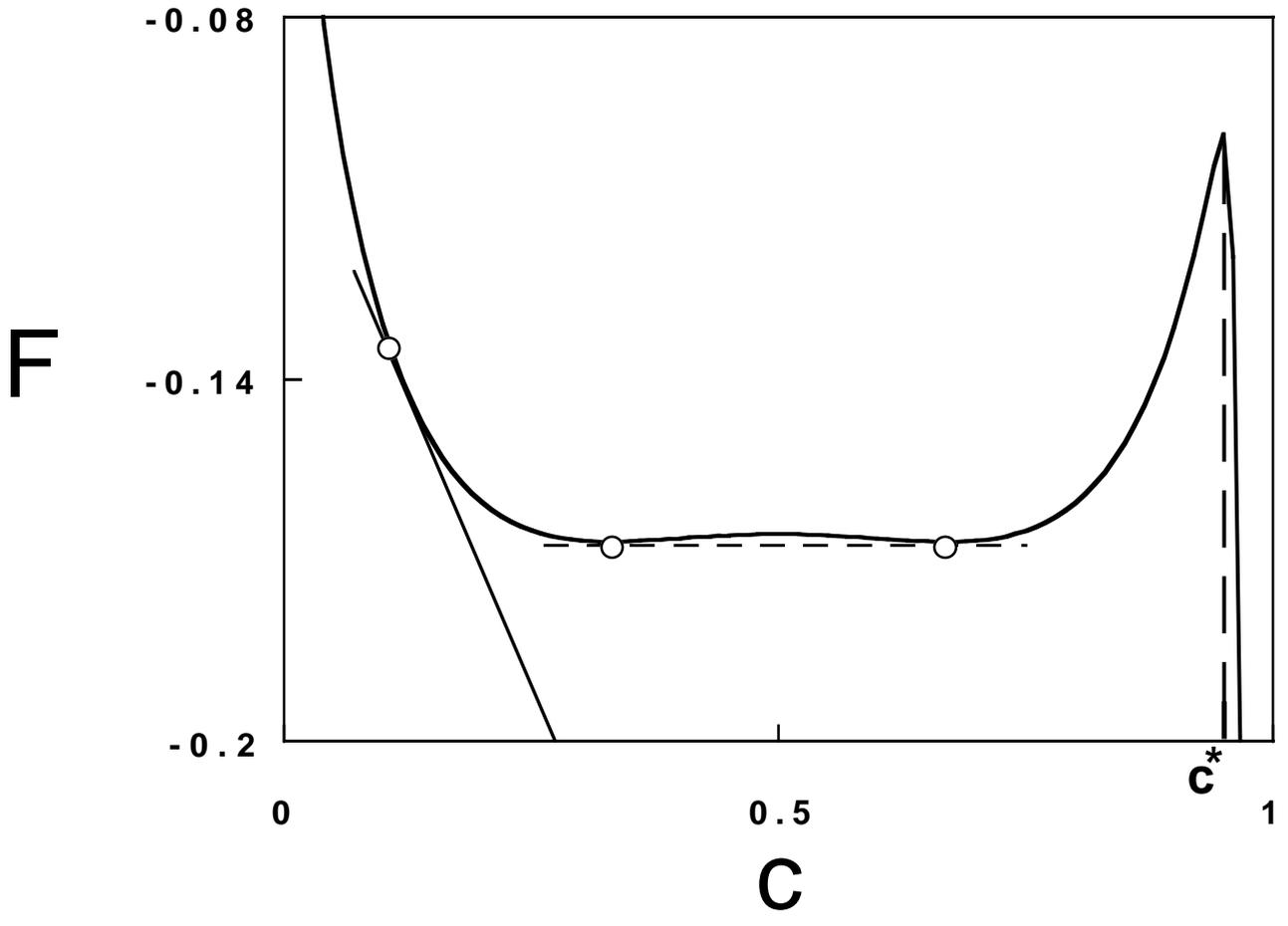
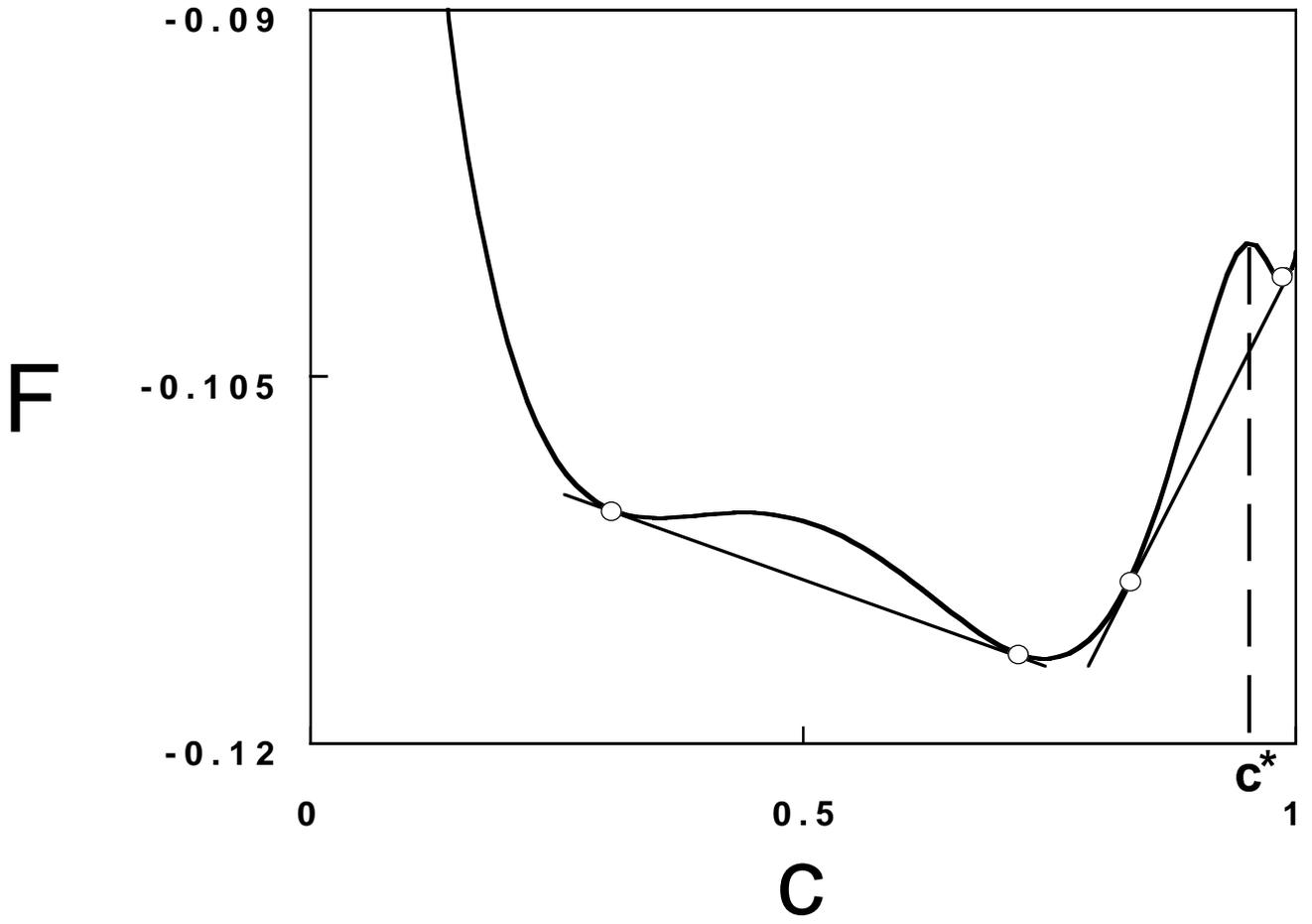
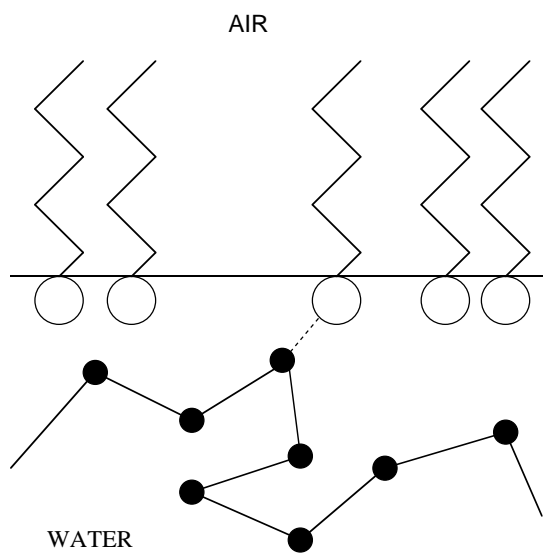
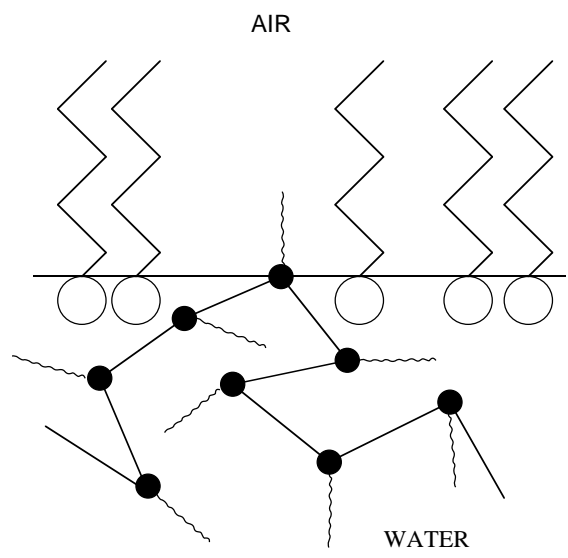
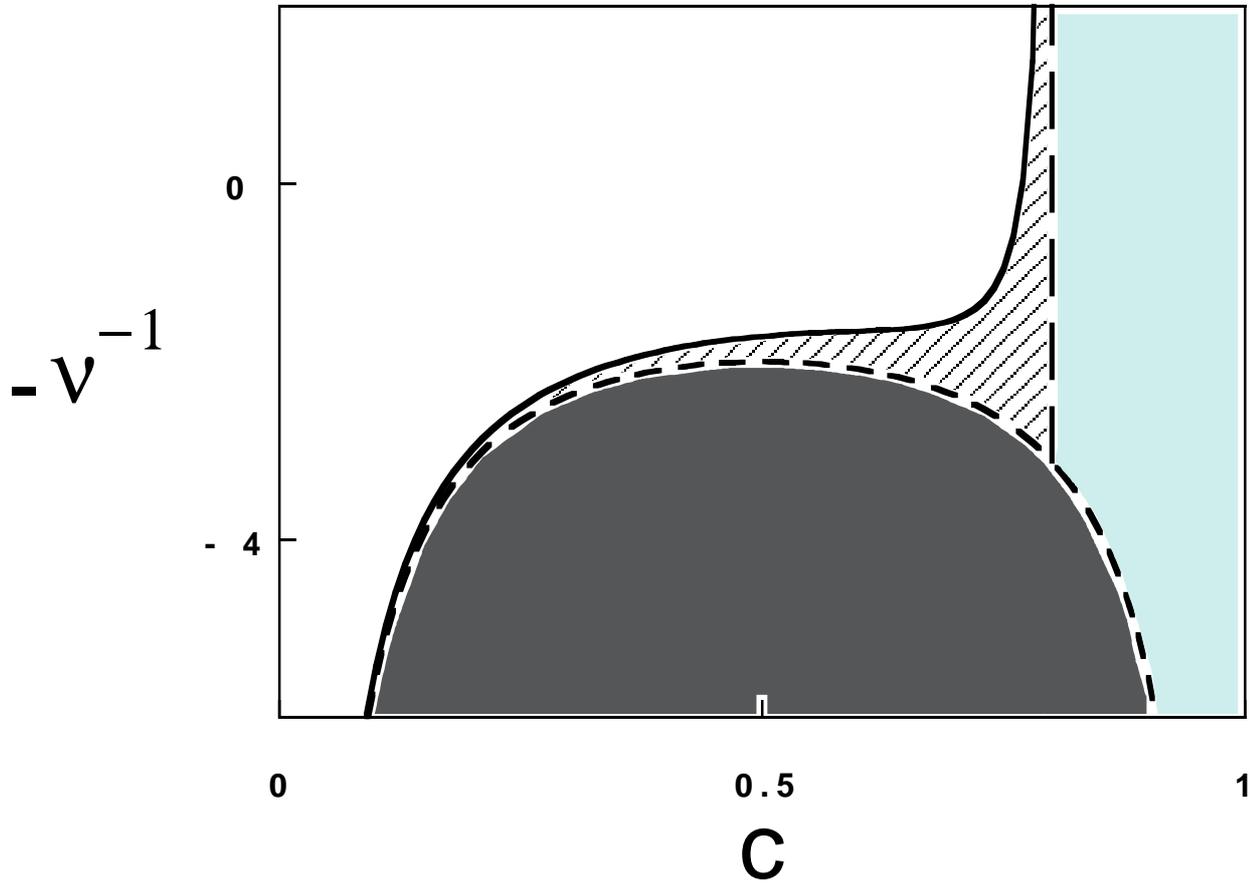
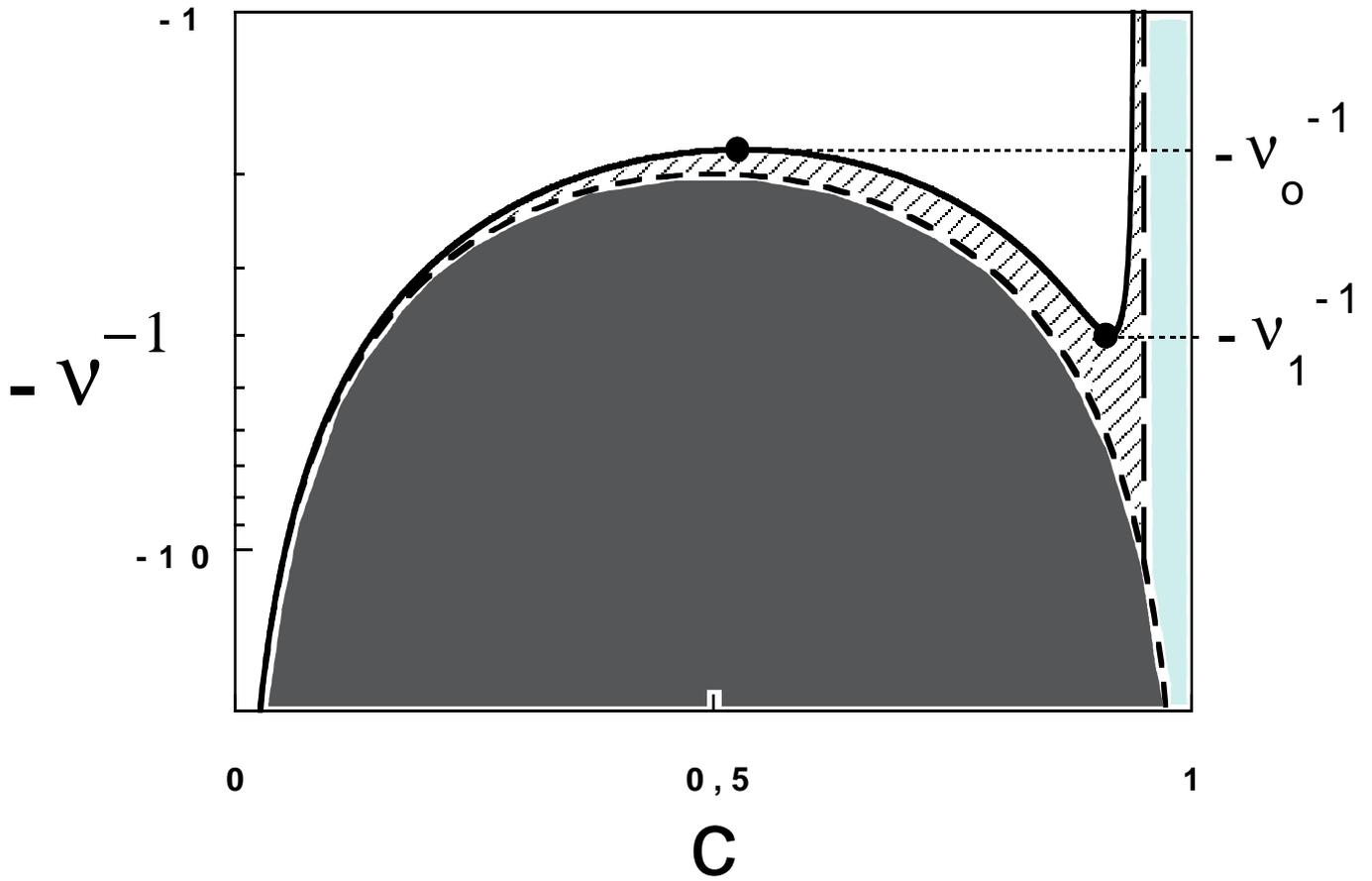
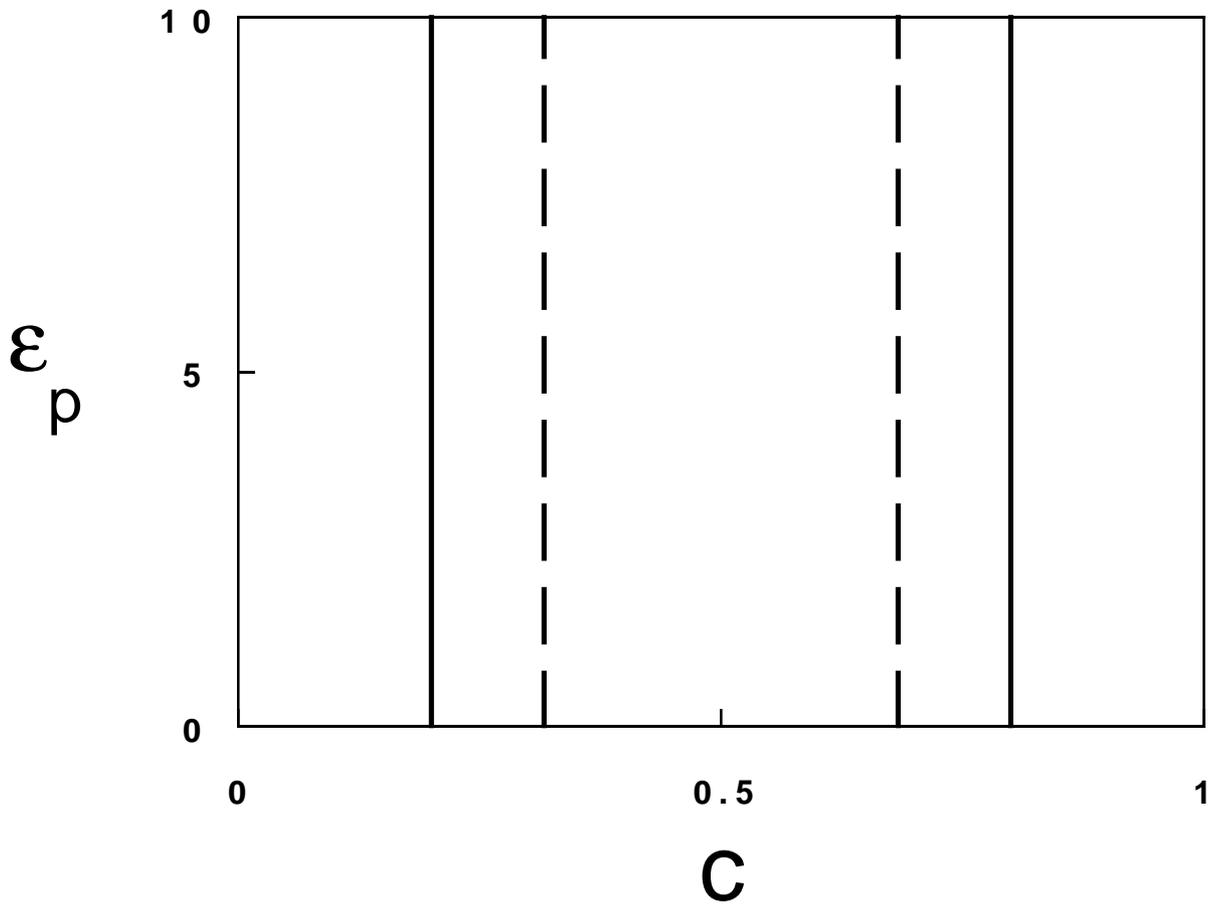
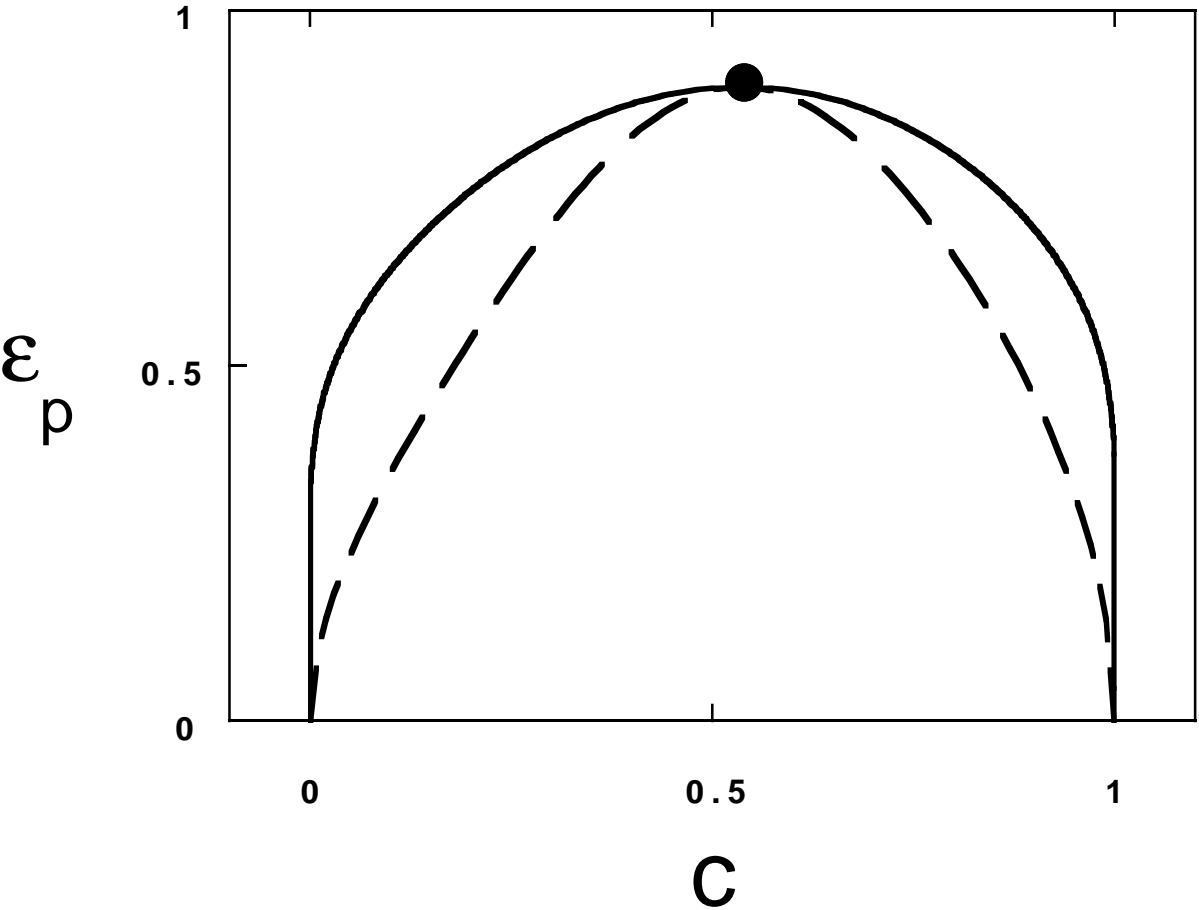
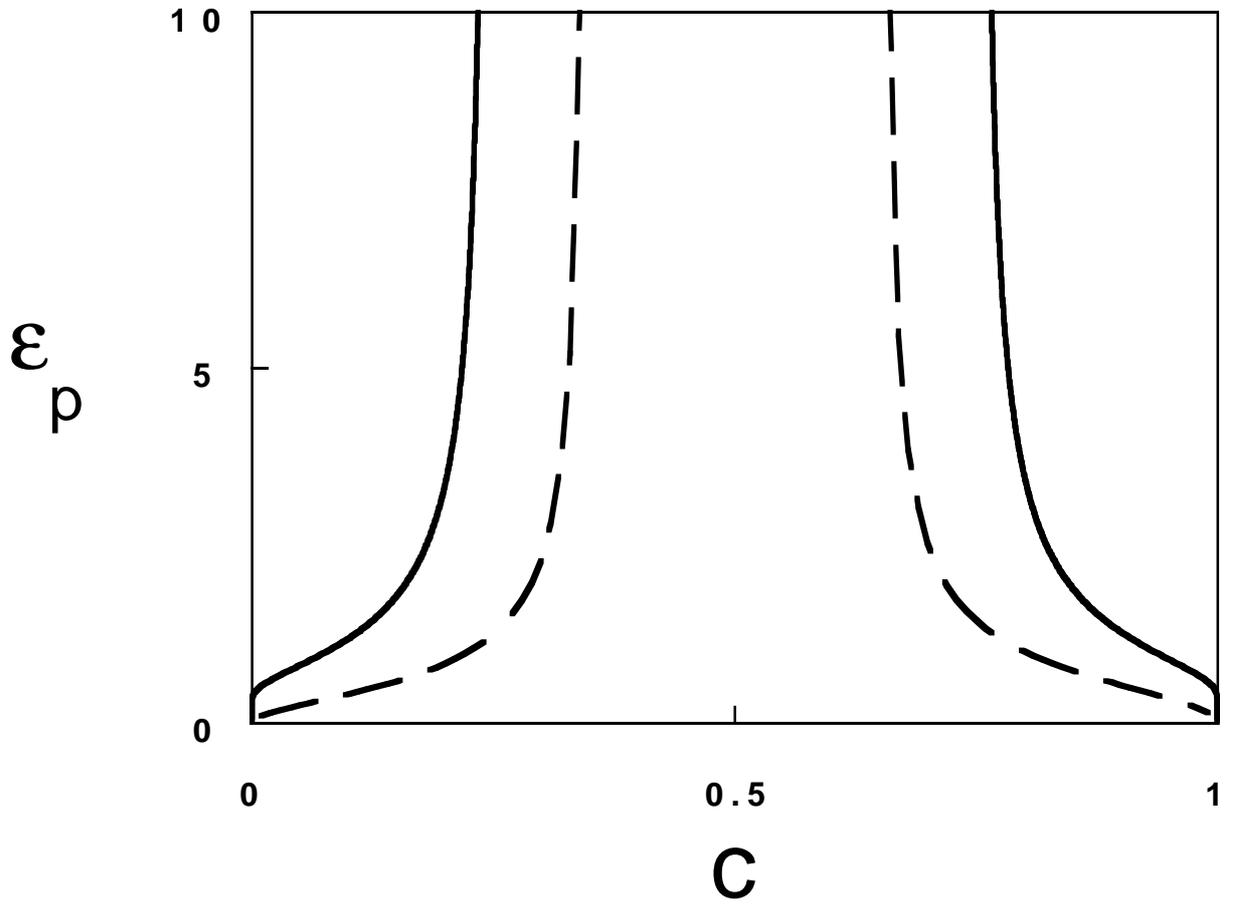
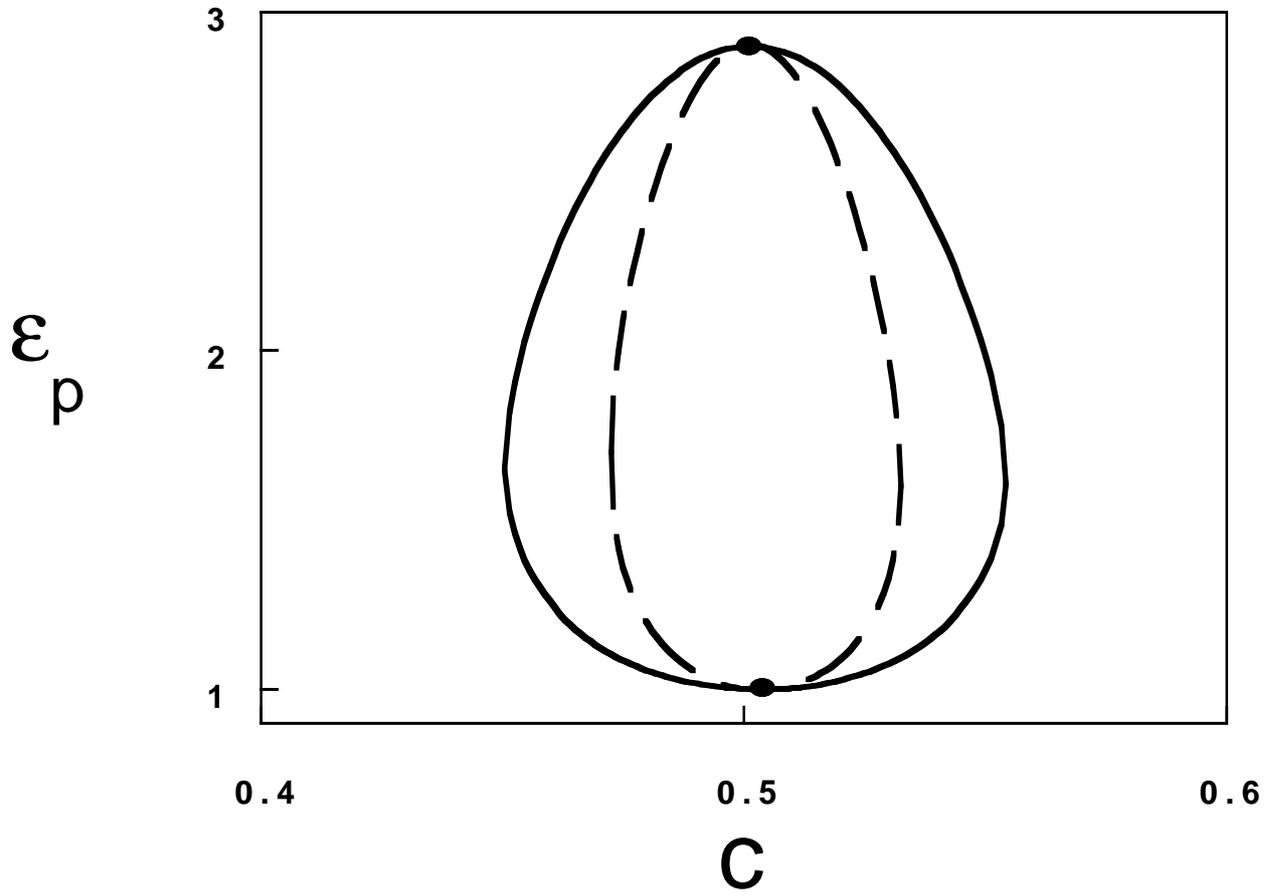
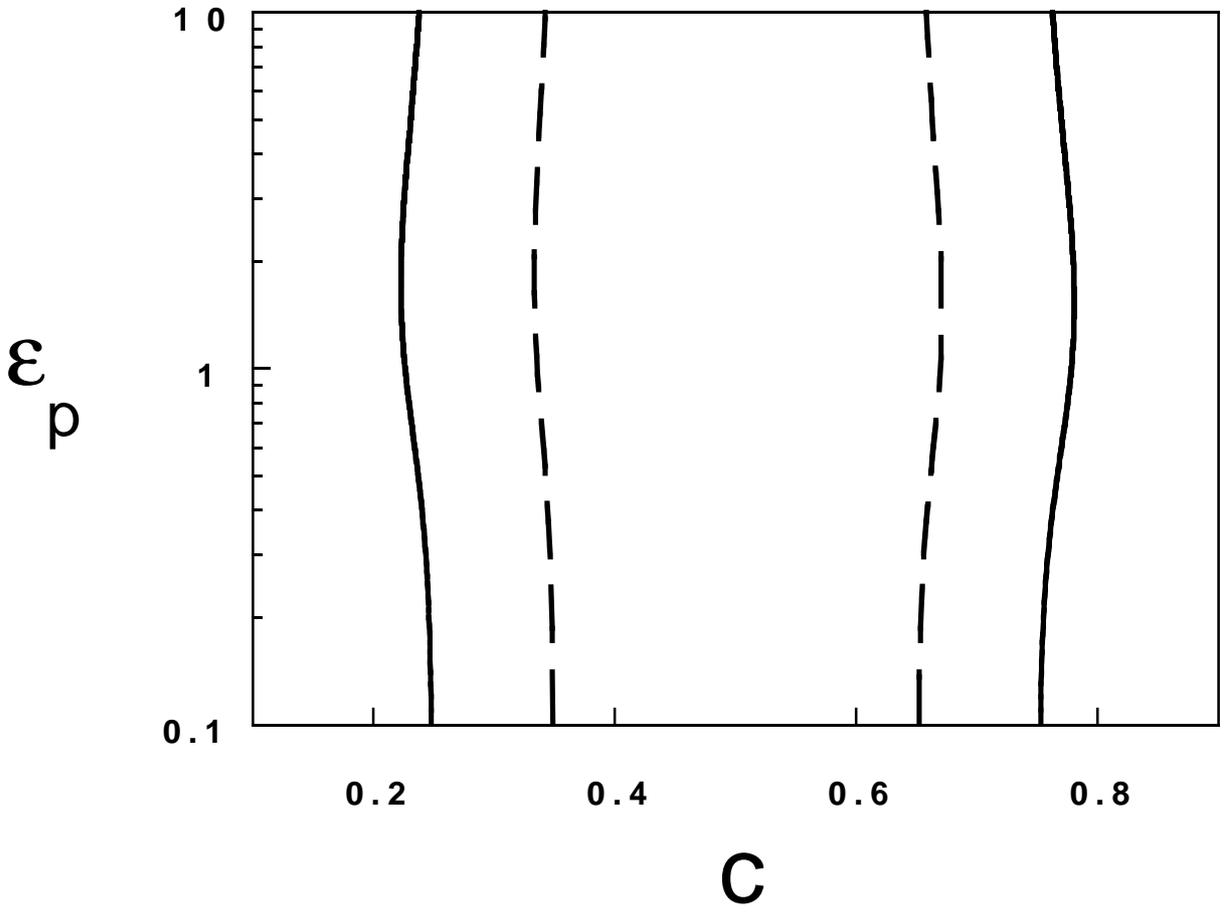
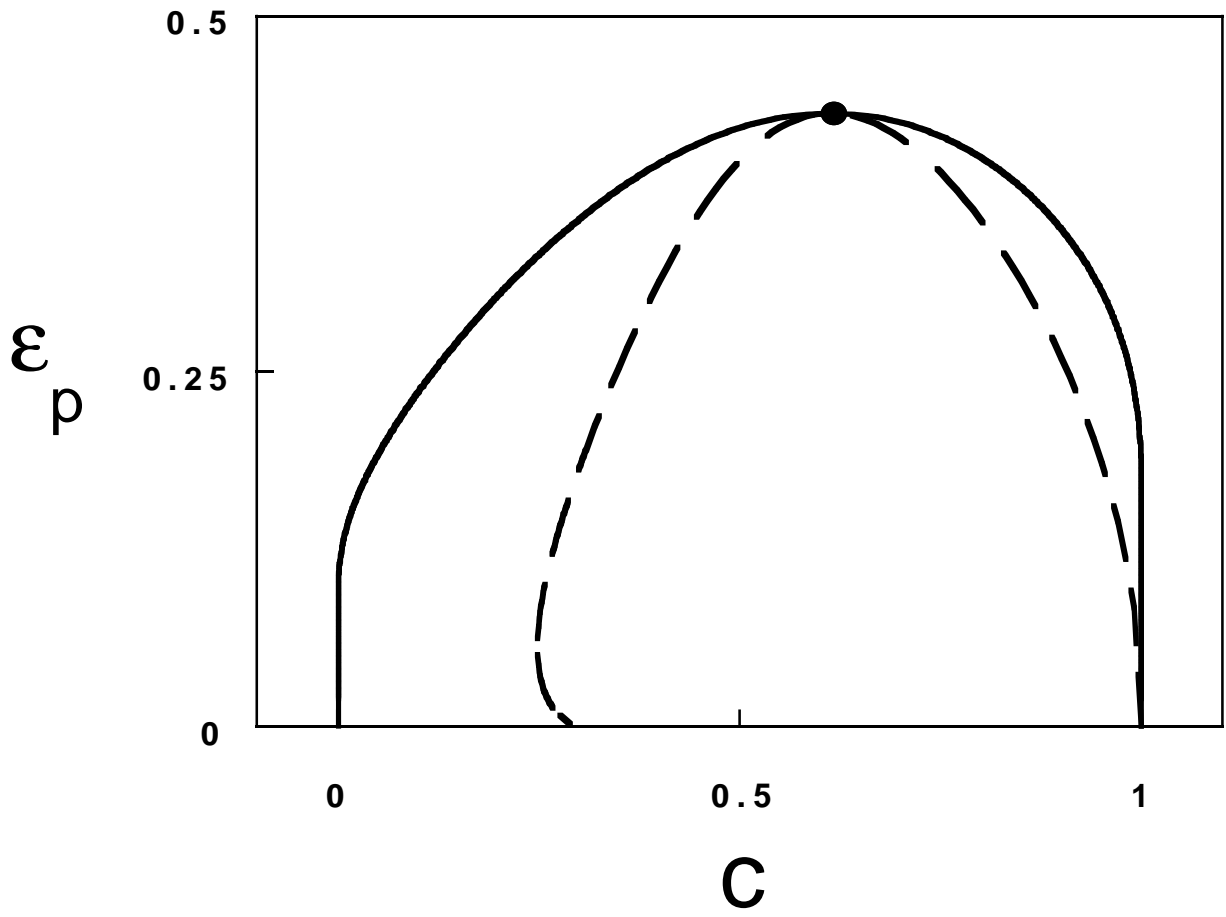
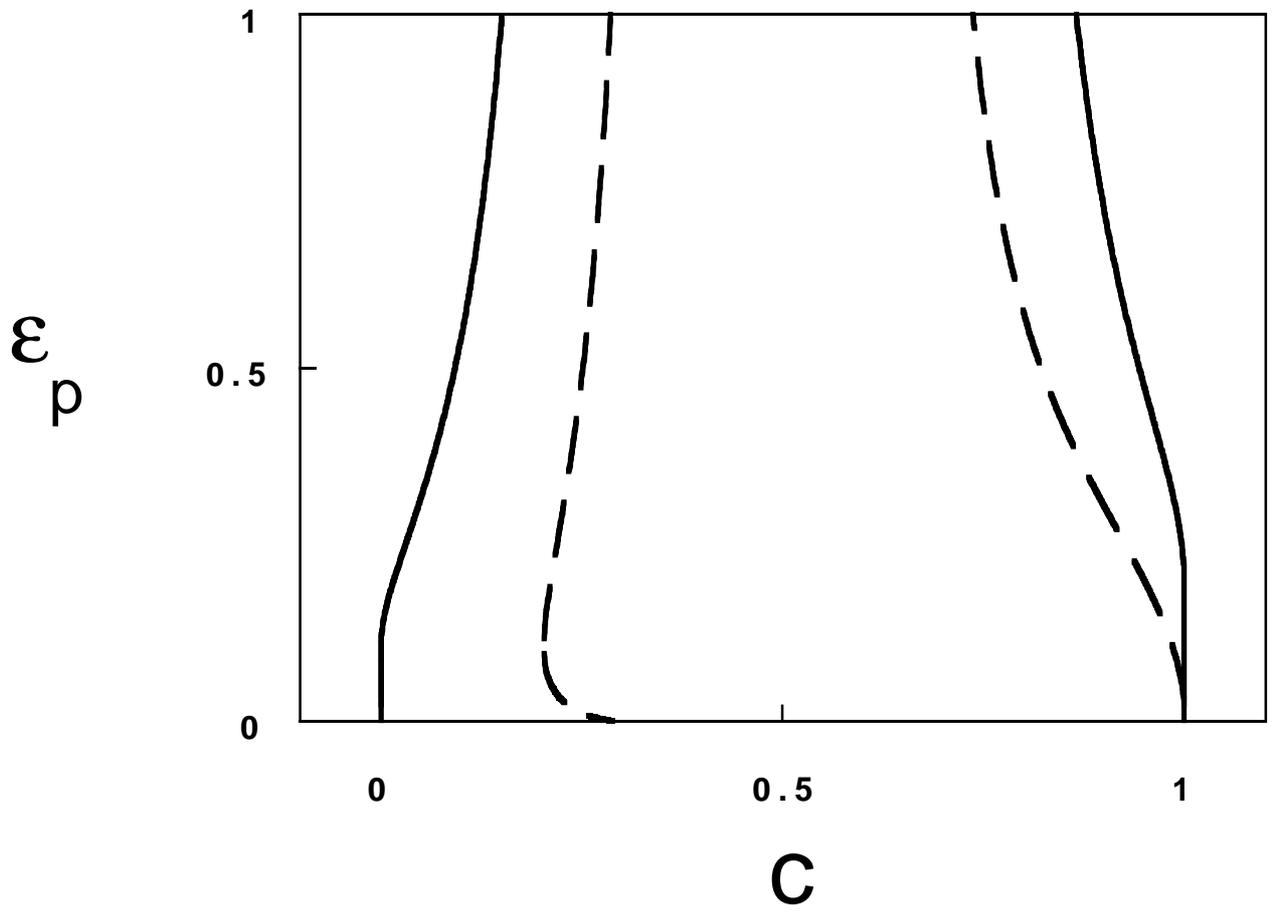
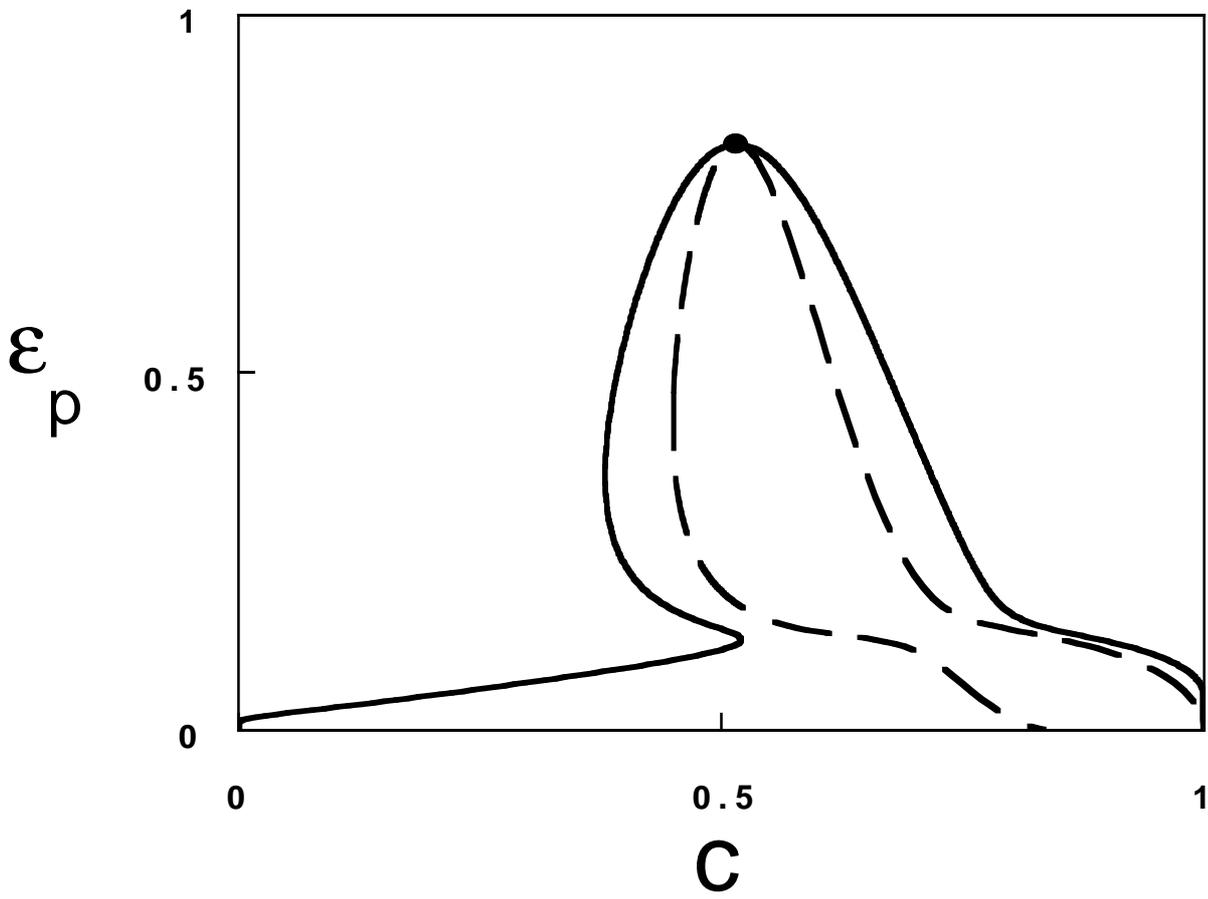
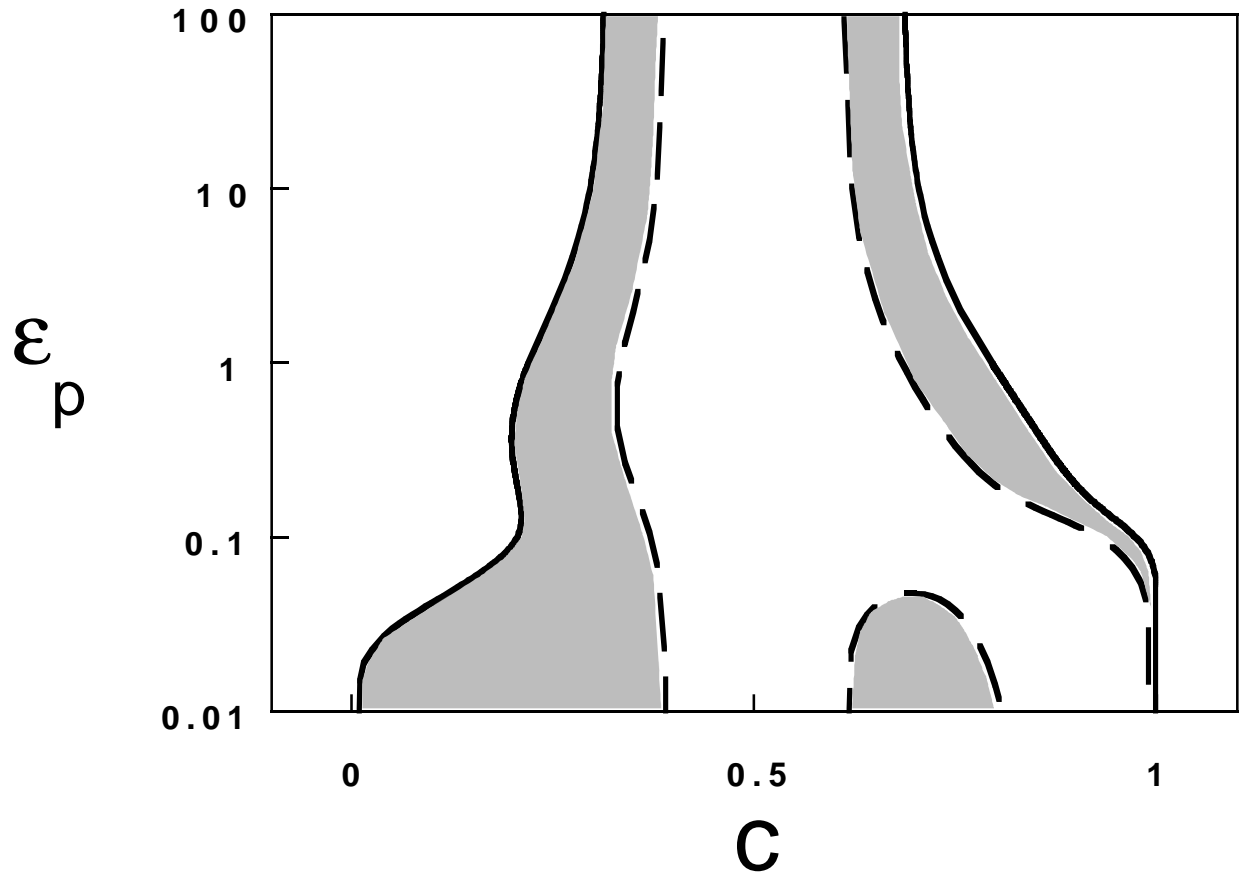
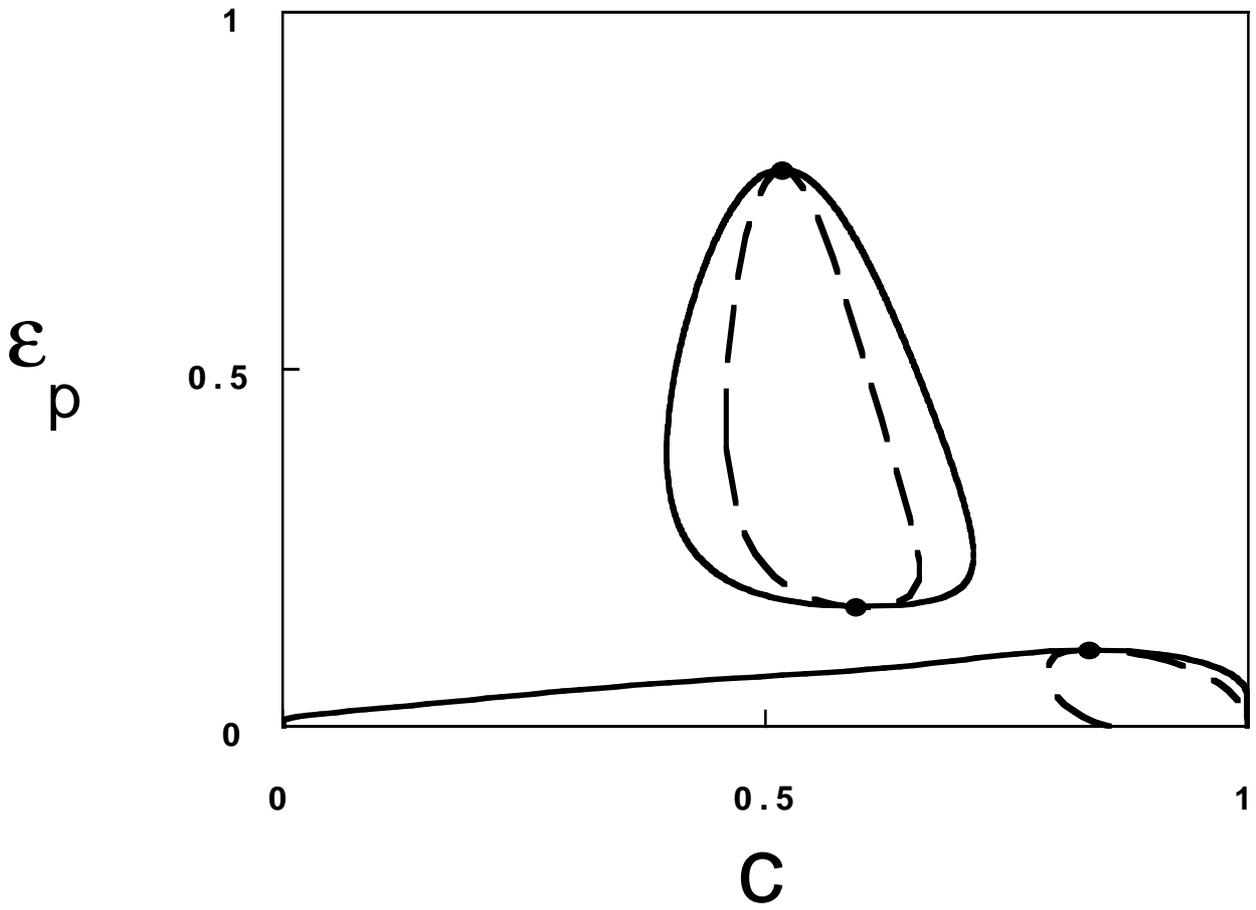
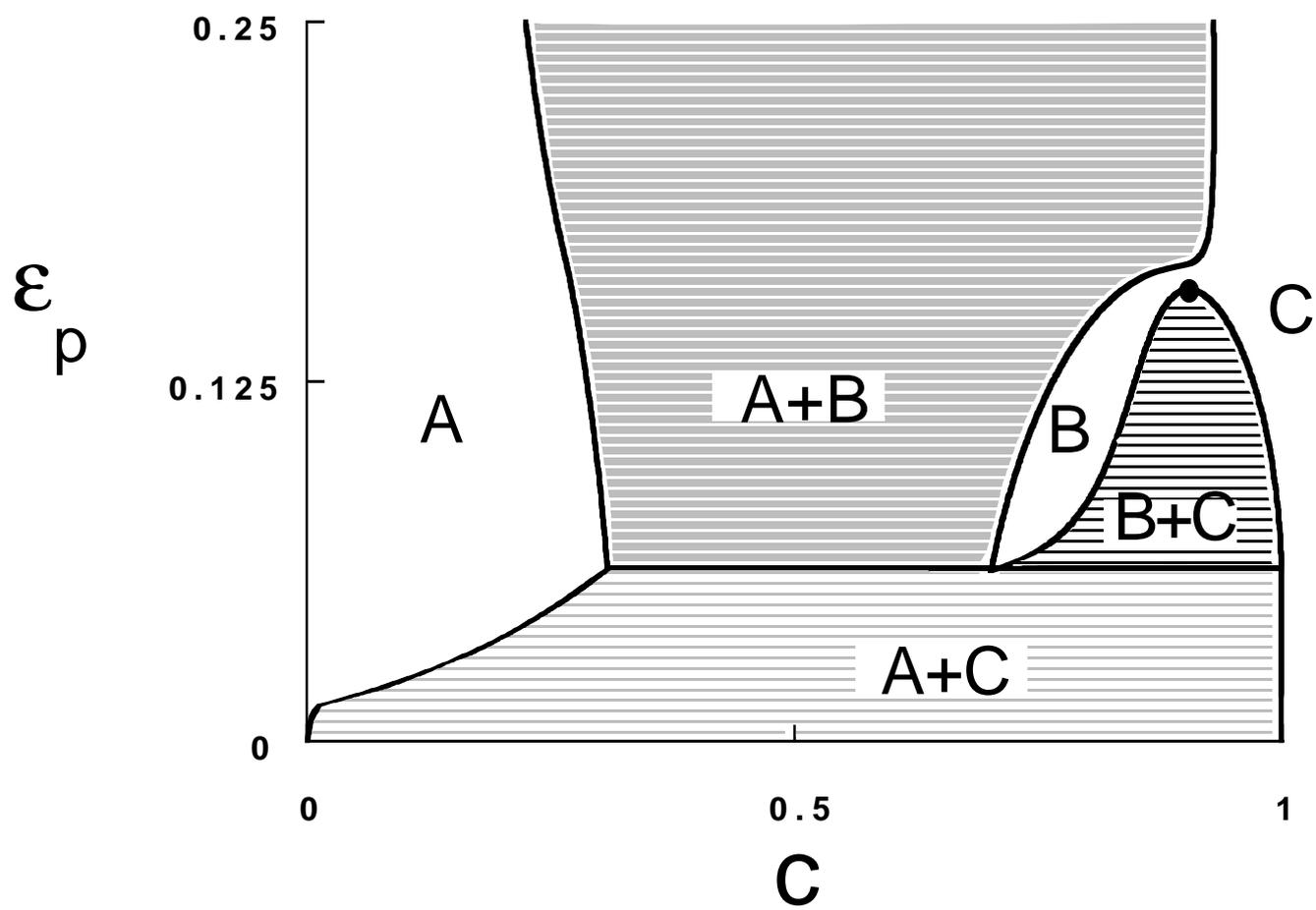
\end{document}